\documentclass[sigplan,10pt,nonacm]{acmart}
\renewcommand\footnotetextcopyrightpermission[1]{}

\usepackage{tikz}
\usepackage{xcolor}

\usepackage[edges]{forest}
\usepackage{listings}
\usepackage[most]{tcolorbox}
\usepackage{balance}

\newtcblisting[auto counter]{apilisting}[2][]{sharp corners, 
    fonttitle=\bfseries, colframe=black, listing only, 
    listing options={basicstyle=\ttfamily,language=c}, 
    title=Listing \thetcbcounter: #2, #1}
    
\definecolor{codegreen}{rgb}{0,0.6,0}
\definecolor{codegray}{rgb}{0.5,0.5,0.5}
\definecolor{codepurple}{rgb}{0.58,0,0.82}
\definecolor{backcolour}{rgb}{0.95,0.95,0.92}

\lstdefinestyle{apistyle}{
    backgroundcolor=\color{white},
    commentstyle=\color{codegreen},
    keywordstyle=\color{magenta},
    numberstyle=\tiny\color{codegray},
    stringstyle=\color{codepurple},
    basicstyle=\ttfamily\scriptsize,
    breakatwhitespace=false,         
    breaklines=true,                 
    captionpos=b,                    
    keepspaces=true,                 
    numbers=none,                    
    numbersep=3pt,                  
    showspaces=false,                
    showstringspaces=false,
    showtabs=false,                  
    tabsize=2,
    frame=lines,	
    escapeinside={\%*}{*)}
}

\usepackage{filecontents}

\newcommand{\stitle}[1]{\vspace{1ex}\noindent\textbf{#1}}
\usepackage{xspace}
\newcommand{\ours}{\textsc{CryptDough}\xspace}
\usepackage{url}

\usepackage[font=small]{caption}
\usepackage[font=footnotesize]{subcaption}

\usepackage{multirow}
\usepackage{graphicx}
\usepackage{comment}

\theoremstyle{definition}

\usepackage{scalerel}

\usepackage{stmaryrd}
\usepackage{MnSymbol}
\usepackage[ruled, linesnumbered]{algorithm2e}
\SetKwInOut{Input}{Input}
\SetKwInOut{vars}{vars}
\SetKwInOut{Output}{Output}
\SetKwInOut{Result}{Result}
\SetKwInOut{Parameter}{Param.}
\SetKwInput{Procedure}{Procedure}
\SetKwComment{Comment}{{//}}{}

\newcounter{protocol}
\newcounter{algorithm}

\IncMargin{0.2em}
\makeatletter
\usepackage{etoolbox}
\patchcmd{\@algocf@start}
  {-1.5em}
  {-0.6em}
  {}{}
  \makeatother

\newcommand{\camera}{}

\usepackage{enumitem}
\setlist{noitemsep,topsep=0pt,parsep=0pt,partopsep=0pt, leftmargin=*}

\begin{document}

\title{\ours: A Unified Analytics Engine for Secure Multiparty Computation}

\settopmatter{authorsperrow=4}

\author{Muhammad Faisal}
\affiliation{%
  \institution{Boston University}
  \city{}
  \country{}}
\email{mfaisal@bu.edu}

\author{Alessandra Lanz}
\affiliation{%
  \institution{Boston University}
  \city{} \country{}
  }
\email{alanz@bu.edu}

\author{Sam Buxbaum}
\affiliation{%
  \institution{Boston University}
  \city{}
  \country{}}
\email{sambux@bu.edu}

\author{Adam Godel}
\affiliation{%
  \institution{Boston University}
  \city{}
  \country{}}
\email{agodel@bu.edu}

\author{Vasiliki Kalavri}
\affiliation{%
  \institution{Boston University}
  \city{}
  \country{}}
\email{vkalavri@bu.edu}

\author{Mayank Varia}
\affiliation{%
  \institution{Boston University}
  \city{}
  \country{}}
\email{varia@bu.edu}

\author{John Liagouris}
\affiliation{%
  \institution{Boston University}
  \city{}
  \country{}}
\email{liagos@bu.edu}

\renewcommand{\shortauthors}{Faisal, Lanz,  Buxbaum, Godel, Kalavri, Varia and Liagouris}

\begin{abstract}

We present \ours, a unified analytics engine for secure multiparty computation (MPC). \ours enables multiple distrusting parties to jointly execute a data analysis pipeline on their private inputs and learn nothing beyond the result (e.g., aggregate statistics). Unlike existing MPC solutions that support a single threat model or workload type, \ours provides built-in support for cross-domain analytics (relational, time series, ML inference) under various threat models, all within the same system runtime. 

\ours contributes (i) a hierarchical system design that facilitates modularity and extensibility through progressive lowering of abstractions, and (ii) the concept of \emph{virtual vectors} that enable users to write single-threaded code across all layers of the software stack, while pushing the complexity of communication, parallelization, and memory management down to the execution engine. We show that \ours generalizes the functionality of state-of-the-art MPC systems and remains competitive on the analytics they support, often outperforming them by more than~$2\times$.

\end{abstract}

\pagestyle{plain}
\keywords{secure analytics; multi-party computation}

\settopmatter{printfolios=true,printacmref=true}

\maketitle

\section{Introduction}
Cryptographically secure multiparty computation (MPC)~\cite{10.1145/3387108} is based on a deceptively simple yet powerful idea: multiple parties can work together to compute the result of a function on their private data without exposing the data to each other or other external entities. The need for such collaborative analyses emerges in various scenarios, where the societal or monetary benefit is magnified if multiple, possibly distrusting, entities allow certain computations on their proprietary datasets while maintaining control over the process.

This work is motivated by three key observations about emerging MPC applications~\cite{mpc-use-cases}. First, \textbf{threat models and security requirements vary widely across use cases}. Some applications require protection against malicious entities~\cite{prio} while others assume ``honest-but-curious'' parties~\cite{student-taxes}.  Second, \textbf{data analysis pipelines are complex and typically include diverse workloads}. The input to many ML tasks is often the result of a relational query~\cite{DBLP:journals/cj/ArcherBLKNPSW18}, while mobile health analytics involve time series operations (windows) followed by relational aggregations~\cite{faisal2023tva}. Finally, \textbf{secure data workflows should be easy to develop without cryptographic expertise}. 
While cryptographers typically work at the level of arithmetic and boolean functionalities, analysts prefer to write programs in high-level languages that must be agnostic to the underlying secure primitives.

This paper presents \ours, \camera{the first} unified analytics system for MPC that meets the above requirements without compromising security or performance.
\ours advances the state-of-the-art in two ways: first, it provides built-in support for mixed data workflows under various threat models and, second, it prioritizes modularity and extensibility at all layers of the software stack. This way, \ours can accommodate a diverse user base (data analysts, software developers, applied cryptographers), which is necessary for democratizing MPC and for keeping up with the rapidly evolving field of cryptographic computing. 

\subsection{Limitations of existing systems}

\ours addresses important limitations of prior work:

\stitle{No built-in support for mixed analytics.}
Existing MPC systems for data analytics (\S\ref{sec:related}) 
target a single workload type, e.g., machine learning, relational queries, or time series computations. The notable absence of holistic approaches presents significant barriers to practical deployments, as stakeholders are typically reluctant to invest in solutions tailored to one application domain. 
For many real use cases that involve mixed workloads, developers are left with two options: either use specialized systems for different parts of the pipeline and do the orchestration manually, or implement the entire pipeline from scratch using a general-purpose compiler framework~\cite{keller2020mp}. Both options require considerable engineering effort and cryptographic expertise. 

\stitle{Non-extensible system designs.}
The vast majority of MPC-based systems for analytics are tailored to specific protocols. This specialization is the result of tight coupling between the target setting (e.g., 3-party computation~\cite{liagouris2023secrecy}) and the system implementation. For instance, protocol-specific communication patterns are often deeply embedded into the runtime, and adapting the respective systems to alternative settings requires a complete reimplementation of core components. While there exist frameworks that support multiple threat models~\cite{cerebro, baum2025orq,watson2022piranha,harth2025pigeon,faisal2023tva}, these systems are designed primarily for analysts and lack high-level abstractions for protocol and system developers to facilitate extensibility.

\stitle{Limited interoperability.} 
One might consider addressing the above limitations by composing mixed pipelines from existing specialized systems. This approach, however, presents technical challenges related to both security and performance. Even if two systems target the same threat model, they often employ fundamentally different data encodings or cryptographic schemes~\cite{shamir, yao-garbled-circuits}. 
Securely converting between representations is non-trivial, and existing tools do not provide standardized interfaces.
Security aside, orchestrating multiple systems is itself a laborious task that requires integrating different runtime environments, communication libraries, data formats, programming APIs, and (often incompatible) execution models. 
We argue that similar limitations of plaintext systems gave birth to unified dataflow engines in the 2010s~\cite{apache-spark,naiad,google-dataflow, dryadlinq}. A decade later, the field of secure computation has reached the same inflection point.

\subsection{Technical challenges and contributions}

\camera{\ours is a system that enables something fundamentally new: executing complex analytical pipelines on secret data owned by multiple parties.  The key challenge lies in architecting a software stack that is modular, easily extensible to new threat models and workloads, and achieves competitive performance without deteriorating into a ``wrapper'' of external black-box libraries.  \ours addresses this challenge by delineating software layer boundaries, defining which primitives and operators belong to each layer, and which abstractions each layer must expose to ensure composability.
While prior work has identified vector abstractions as necessary for good MPC performance~\cite{liagouris2023secrecy,baum2025orq}, none show how such abstractions can be composed to reuse low-level primitives in high-level secure operators. \ours enables exactly this transparent composition~and~reuse.}

\noindent\stitle{Contributions.} We make the following contributions:
\begin{itemize}
  \item We identify a minimal set of secure MPC functionalities ($+, \times, \oplus$, $\land$, etc.) that are sufficient to express all workload types we consider in this work. 
\ours exposes a vectorized interface and utilities for protocol developers to implement efficient versions of these functionalities. 
  \item We design and implement a small set of secure, protocol-agnostic   primitives (e.g., comparisons, multiplexers, etc.) that treat low-level MPC functionalities as black boxes. \ours allows composing these primitives into high-level operators to ensure that user programs inherit the security guarantees of the underlying functionalities.
  \item  We introduce \emph{virtual vectors}, a mechanism that enables \ours users to write single-threaded code across all layers of the stack and push the complexity of communication, parallelization, and memory management to the system runtime. With virtual vectors, 
software developers can reuse the vectorized primitives to implement secure, data-parallel operators with irregular access patterns. 
  \item We use \ours's abstractions to implement four MPC protocols across the entire spectrum of threat models, and three secure analytics libraries (relational, time series, ML inference) that can be used to create mixed data pipelines. We show that \ours generalizes the functionality of specialized systems and often exceeds their performance for the workloads they support.
\end{itemize}\vspace{1mm}

\noindent
\camera{The \ours source code is available at \url{https://github.com/CASP-Systems-BU/CryptDough}.}
\section{\ours overview}\label{sec:overview}

\begin{figure}[t]
  \centering
      \includegraphics[width=\linewidth]{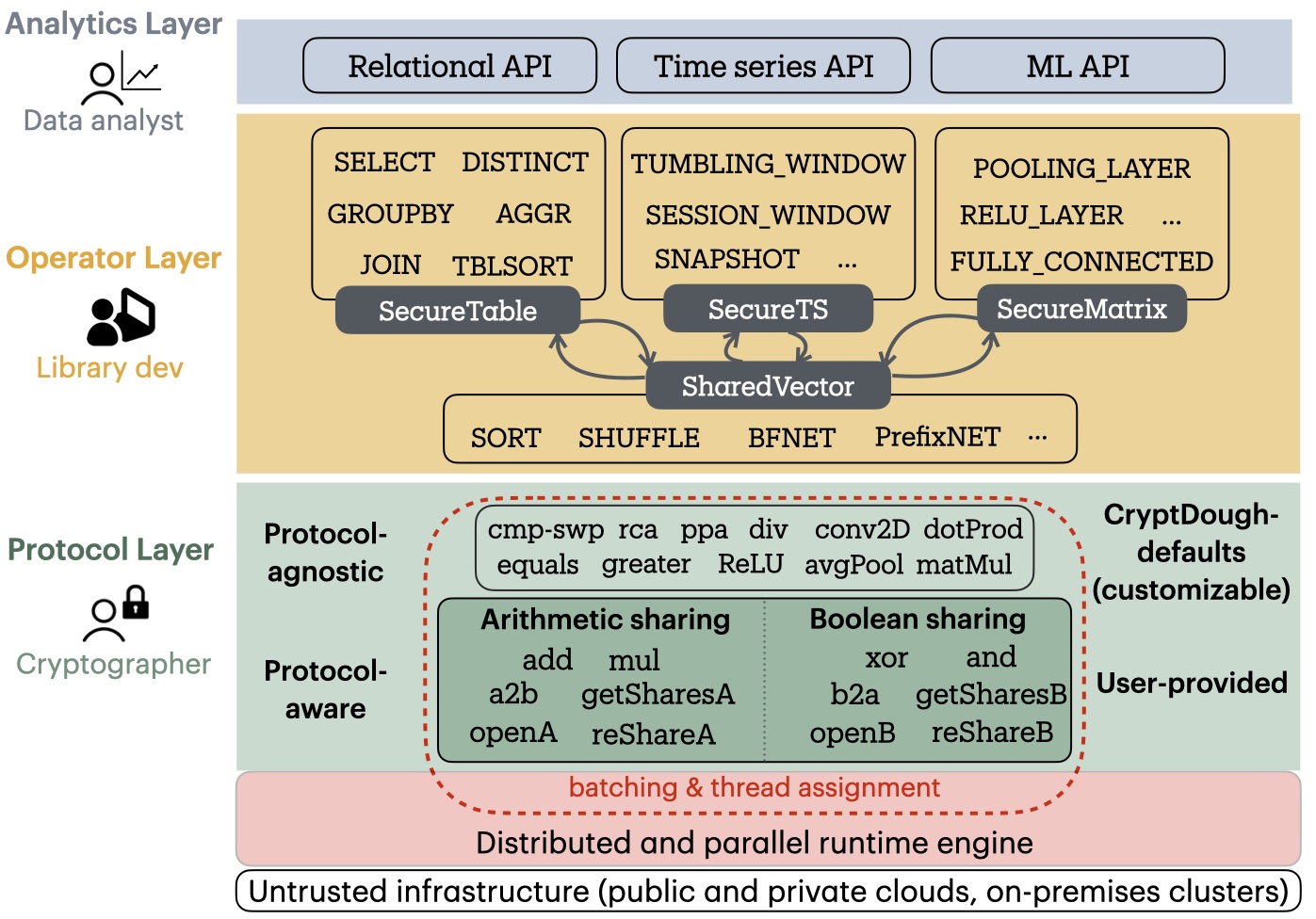}\vspace{-2mm}
  \caption{\ours's software stack and target users}
  \label{fig:stack}
\end{figure}

\ours is an elaborate software stack in which all parts are carefully crafted to support each other, prioritizing modularity, extensibility, and separation of concerns. We have designed \ours with a hierarchy of the form: \emph{Low-level MPC Functionalities} $\rightarrow$ \emph{Oblivious Primitives}$\rightarrow$ \emph{High-level
Operators} $\rightarrow$ \emph{Operator Composition}. All layers of the stack, except the bottom one, are protocol-agnostic and each layer makes black-box use of primitives from the layers below within a \textbf{fully oblivious control flow}. In practice, this means that the \camera{source code} contains no conditional branches that depend (directly or indirectly) on secret data, and that user programs execute an identical sequence of operations for all inputs of the same size. Fig.~\ref{fig:stack} shows \ours's \camera{layered} architecture \camera{and target users}.

\subsection{Design principles}
\ours's architecture relies on three core principles: \vspace{1mm}

\begin{enumerate}
	\item \textbf{Configurable security.} MPC protocols have diverse trust assumptions, 
and no single threat model is suitable for all scenarios. 
For example, internal enterprise MPC deployments can assume semi-honest parties, as employees are bound by legal agreements that 
discourage deviations from the protocol.  On the other hand, cross-organization deployments may require protection from malicious adversaries. 
\ours accommodates this diversity by supporting MPC protocols with varying threat models. 

	\item \textbf{Extensibility.} The system must be easy to extend along two dimensions: (i) new MPC protocols, and (ii) new domain-specific libraries for analytics. Extensibility is achieved via \ours's layered architecture and well-defined interfaces that isolate protocol-level concerns from application logic, enabling independent evolution of both cryptographic primitives and analytics capabilities.
	
	\item \textbf{Practical performance.} Generality must not come at the expense of performance. As we show in \S\ref{sec:evaluation}, \ours is competitive with secure analytics systems that are purpose-built for specific workloads or protocols. We achieve this through a combination of efficient runtime execution strategies, optimized communication primitives, and the ability for expert users to provide custom implementations when necessary. Importantly, \ours never trades security for performance.
\end{enumerate}
	
\subsection{Target users}\label{sec:users}
\ours's layered architecture enables three fundamentally different types of users, each with orthogonal skills and objectives, to contribute to and benefit from the system independently, without requiring cross-domain knowledge. Specifically, \ours targets the following user profiles:

\stitle{A data analyst} is the end-user, e.g., a data scientist, who builds data analysis pipelines. 
This user has domain expertise in analytics, but no cryptographic background and no prior knowledge of MPC. 
\ours enables data analysts to compose secure dataflows using a familiar declarative DSL, similar to those offered by mainstream plaintext systems like Apache Spark~\cite{apache-spark} and Apache Flink~\cite{carbone2015apache}. The DSL abstracts away all cryptographic and system-level complexity, allowing data analysts to focus on the logic of their programs.

\stitle{A library developer} is a domain expert in a specific analytics workload (e.g., ML, relational analytics, genomics, etc.) who wishes to extend \ours with secure operators for their domain. This user knows how to design secure operators with data-independent control flows, but is not familiar with the internals of cryptographic primitives and may have limited experience \camera{with system development and optimization}. 
\ours enables library developers to implement domain-specific operators that are compatible with all supported protocols at the lowest layer.

\stitle{An applied cryptographer} is a user who wishes to extend \ours with new MPC protocols. 
While possessing deep cryptographic expertise, the applied cryptographer may lack systems programming skills, parallelization knowledge, or data analytics background. \ours allows such users to contribute new protocols by specifying only a minimal set of low-level functionalities, such as secure addition, multiplication, and boolean operations. To further simplify protocol implementation, \ours also  provides access to a communicator and a randomness generation module.\vspace{1mm}

\noindent A key benefit of \ours's architecture is that \emph{contributions from each user type immediately benefit all others} without additional integration effort. For example, new graph analytics operators contributed by a library developer would be immediately compatible with all existing MPC protocols in \ours, inheriting their security guarantees and threat models. Similarly, when an applied cryptographer adds a new protocol, all existing analytics libraries can be directly used under its threat model without modification.  

While \ours defines clear abstractions to isolate users' concerns, it also provides flexibility for advanced \emph{customization} (Fig.~\ref{fig:stack}, \emph{\ours-defaults}).  An applied cryptographer with programming expertise can bypass default parallel implementations and provide protocol-optimized versions of higher-level functionalities for improved performance (we give an example in \S\ref{sec:intermediate-primitives}).  \ours also allows library developers to implement custom primitives at lower layers of the software stack to exploit domain-specific optimization opportunities. 

\subsection{Supported workloads}\label{sec:workloads}
\camera{Real-world MPC deployments today are few and consist mainly of simple data processing, statistical analysis, secure aggregation, and basic ML~\cite{mpc-use-cases}. \ours's main motivation is to push MPC toward more complex analytics.}

To that end, \ours currently supports three workload types that users can combine into end-to-end secure data pipelines: (i) relational analytics, (ii) time series computations, and (iii) ML inference.
\camera{We chose these workloads specifically because they are sufficient to express real-world use cases we have explored in collaboration with stakeholders: a mobile health application that requires secure predictions on time series data, and an education study that requires ML predictions on relational data such as exam scores, attendance records, and family histories.}

\stitle{Relational analytics.} \ours provides APIs for all common relational operators and allows their arbitrary composition into secure query plans. Specifically, it offers the same functionality as state-of-the-art relational MPC frameworks~\cite{baum2025orq,Poddar2021Senate,fang2024secretflow}, including ORQ's shuffling and sorting protocols and join-aggregation optimizations for efficiently evaluating acyclic multi-join queries. We have further extended this functionality with (i) fixed-point arithmetic, (ii) optimized prefix networks for efficient aggregations, and (iii) alternative sorting networks, such as bitonic sort.

\stitle{Time series analytics.} The time series library subsumes the functionality of prior works in this area~\cite{faisal2023tva,waldo}. \ours supports temporal operators that compute aggregates over specified time intervals, as well as tumbling and session windows (both gap-based and threshold-based). We have also implemented all time-agnostic operators from TVA and extended them to fixed-point arithmetic.

\stitle{ML inference.}
The ML library supports common inference primitives offered by state-of-the-art systems, such as Piranha~\cite{watson2022piranha} and Pigeon~\cite{harth2025pigeon}, including convolution, ReLU, pooling, and fully-connected layers. These primitives enable the construction of neural networks such as VGG-16 and AlexNet. \ours  also provides secure dot product and matrix multiplication primitives that can be used to implement simpler regression models.

\begin{figure}[t]
  \centering
      \includegraphics[width=\linewidth]{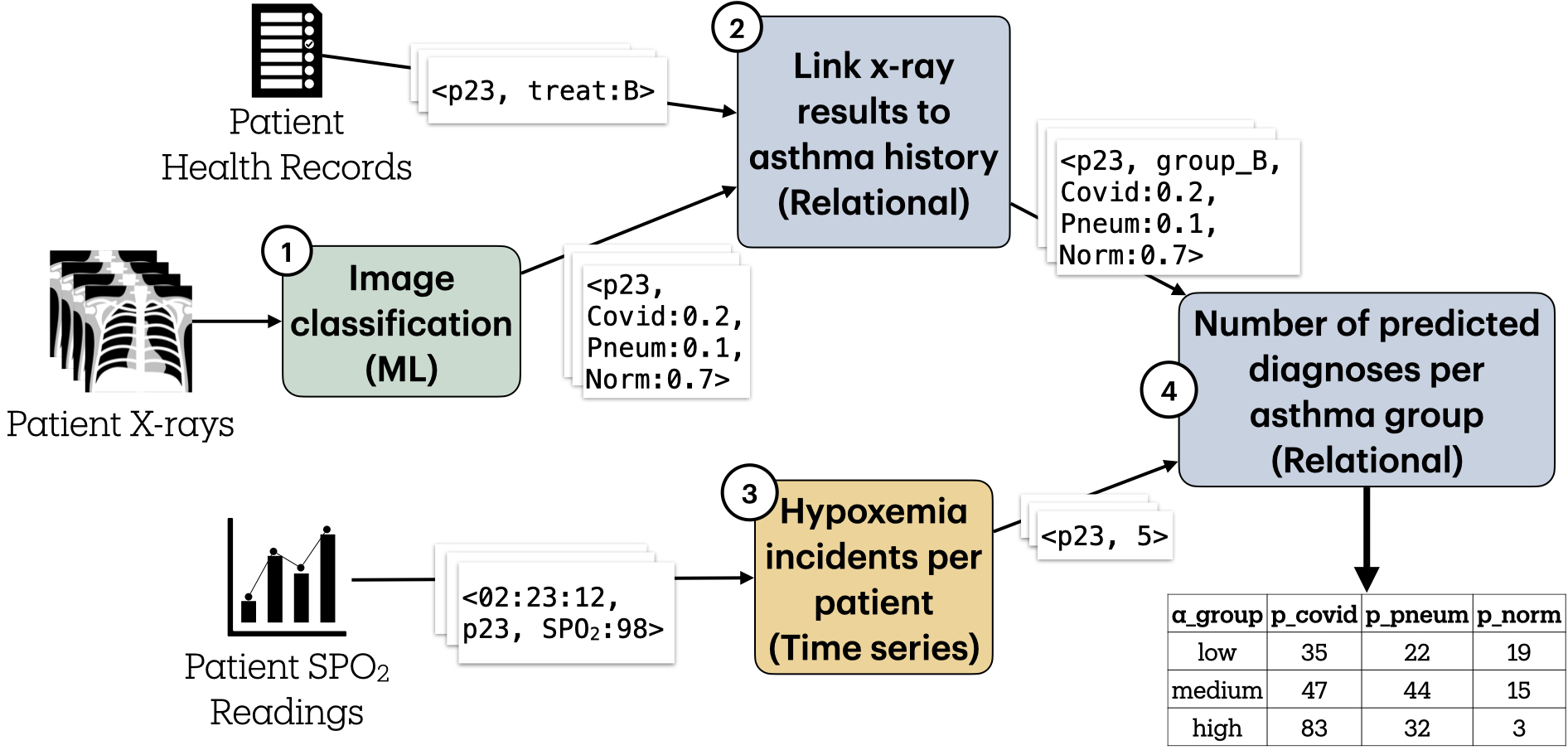}\vspace{-1mm}
  \caption{A mixed workflow example that analyzes diverse sensitive data from hospitals, radiology centers, and wearable devices.}
  \label{fig:mixed-pipeline}
\end{figure}

\stitle{Example workflow.} Fig.~\ref{fig:mixed-pipeline} shows a mixed analytics workflow and \camera{Listing~\ref{lst:multi-workload} shows how data analysts can implement it in \ours.} The pipeline combines image data (X-rays), tabular data (health records), and time series data (oxygen saturation measurements) to identify whether a history of hypoxemia incidents in asthmatic patients may be an indicator for diagnosing pneumonia or COVID-19 from their X-rays. The workflow begins with an image classification task on the X-rays (Step 1) to compute probabilities for three  outcomes: COVID-19,  pneumonia, and normal. Next, the classification output is joined with a table containing asthmatic patient records and their treatment status (Step 2). The third stage computes hypoxemia incidents per patient by analyzing $SPO_2$ measurements using a threshold window operator (Step 3). This is a complex aggregation that first identifies time periods (windows) where a patient's $SPO_2$ falls below a threshold (92) and then counts the number of such windows.  Finally, the workflow selects patients with X-ray predictions greater than 0.8 and at least one hypoxemia incident, and computes the total number of patients per outcome and asthma treatment group (Step 4).

\begin{lstlisting}[language=C++, caption={The workflow of Fig.~\ref{fig:mixed-pipeline} implemented in \ours. We omit data source configuration and some minor data transformations. The program can be compiled against any of the supported MPC protocols at the lowest layer, with no modification.}, numbers=left, label={lst:multi-workload}, escapechar=|]
int main(int argc, char** argv) {
  EngineRef eng = init(argc, argv);	// Engine handler
  // Time series schema ([X] means boolean shares of X)
  auto ts_schema = {"[TS]", "[PATIENT_ID]", "[SPO_2]",
  								  "[INCIDENT_ID]", "[INCIDENT_CNT]"}; 
  // Relational schema
  auto rel_schema = {"[PATIENT_ID]", "[ASTHMA_GROUP]"};
  // Classification output schema
  auto cl_schema = {"[PATIENT_ID]", "P_COVID", 
  									"P_PNEUM", "P_NORMAL"};
  // Load secret shares
  SecureTS<T> spo2 = eng.load(ts_schema, ...);
  SecureTable<T> patients = eng.load(rel_schema, ...);
  SecureMatrix<T> xrays = eng.load(IMG_SIZE, ...);
  // Initialize model (cf. |Appendix~\ref{apdx:vgg16}|) and load weights
  Model vgg16 = VGG16(eng);
  vgg16.load(...);
  // Initialize predictions table
  SecureTable<T> predictions = eng.init_table(cl_schema, 
  																					  ROWS, ...);    
  
  // Step 1: Classify X-rays with pretrained VGG16 model
  SecureMatrix<T> vgg_out = vgg16.forward(xrays);
  // Extract model output and insert into predictions
  vgg_out.as_table(predictions);
  
  // Step 2: Join patients and prediction tables
  auto pat_pred = patients.join(predictions, 
      				 						 {"[PATIENT_ID]"}); // Join key
  // Columns to keep in the join output    				 		
  pat_pred.project(cl_schema);
  
  // Step 3: Count hypoxemia incidents per patient
  spo2.session_window({"[PATIENT_ID]"}, "[SPO_2]", 
  										 "[INCIDENT_ID]", 92);
  spo2.filter(spo2["[INCIDENT_ID]"] > 0);
  spo2.distinct({"[PATIENT_ID]", "[INCIDENT_ID]"});
  spo2.aggregate({"[PATIENT_ID]"}, 
  							 {"INCIDENT_CNT", COUNT});
  // Identify patients with at least one incident
  spo2.filter(spo2["[INCIDENT_CNT]"] > 0);
  
  // Step 4: Join outputs of steps 2 and 3 and aggregate
  auto result = spo2.join(pat_pred, 
      										{"[PATIENT_ID]"});  // Join key
  // Columns to keep in the join output    				 		
  result.project({"[A_GROUP]", "P_COVID", "P_PNEUM", 
  								"P_NORMAL"});		
  // Identify patients with high prediction probability
  result["CNT_COVID"] = result["P_COVID"] > 0.8;
  result["CNT_PNEUM"] = result["P_PNEUM"] > 0.8;
  result["CNT_NORMAL"] = result["P_NORMAL" > 0.8;
  result.aggregate({"[A_GROUP]"},	// Grouping key
        		{	// Count patients per outcome
        			{"CNT_COVID", "CNT_COVID", COUNT},
        			{"CNT_PNEUM", "CNT_PNEUM", COUNT},
        			{"CNT_NORMAL", "CNT_NORMAL", COUNT},
        		});
  auto output = result.open_with_schema();
}

\end{lstlisting}

This workflow would be extremely difficult or impossible to express in existing MPC systems. While individual steps could potentially be implemented with specialized frameworks (e.g., Piranha~\cite{watson2022piranha} for image classification, ORQ~\cite{baum2025orq} for relational operations, TVA~\cite{faisal2023tva} for time series), the user would need to write custom integration code to extract output from one system and ingest it into the next.  \camera{The challenge here is that different systems may target different threat models (semi-honest or malicious), making a combination impossible. But even under the same threat model, data encodings often differ. For example, Pigeon~\cite{harth2025pigeon} and ORQ both support semi-honest, 3-party protocols but use different data sharing schemes, so passing output between them requires understanding and writing custom cryptographic code.}
 
Alternatively, one could implement the entire workflow in a general-purpose MPC compiler, such as MP-SPDZ~\cite{keller2020mp}, but this would require several hundred lines of low-level code \camera{compared to the dataflow program in Listing~\ref{lst:multi-workload}.
Without high-level abstractions, users must manually compose secure primitives (e.g., addition, multiplication, comparison) into secure operators. For each operator they must design an oblivious algorithm, derive padding bounds to prevent leakage, and apply batching and data parallelism to achieve reasonable performance. To get a sense of the effort required, implementing one relational query directly from low-level primitives takes $\sim800$~LOC, even when using \ours's vectorized API that substantially reduces complexity. In an earlier version of our codebase that lacked vectorized abstractions, the same query required about ~$2\times$ more~LOC.} 

\section{Threat models and security guarantees}\label{sec:background}
A typical MPC setting involves three types of entities: (i) \textbf{data holders}, who own the private input data, (ii) \textbf{computing parties}, which execute the MPC protocol on encodings of the input data to compute encodings of the result, 
and (iii) \textbf{output parties}, who learn the output of the computation. In \ours, data holders, computing parties, and output parties are separate entities by default but may have more than one role in general; for example, data holders can also act as computing~parties or analysts.

\stitle{Secret sharing.} \ours protects data throughput the entire computation using secret sharing~\cite{shamir}. A data owner encodes a private value $s$ into random $\ell$-bit \emph{shares} $s_1, s_2, \dots, s_m$ and distributes them across computing parties. No individual share reveals information about the secret, but all together are sufficient to reconstruct it. \ours supports both \textbf{arithmetic} secret sharing, where shares add up to the secret using modular arithmetic (i.e., $s = s_1 + s_2 + \dots + s_m~~\texttt{mod}~~2^\ell$), and \textbf{boolean} secret sharing, where shares XOR to the original secret (i.e., $s = s_1 \oplus s_2 \oplus \dots \oplus s_m$). Some operations on shares are local, while others require communication between parties. For additive sharing, if parties hold shares of $x$ and $y$, they can obtain shares of $x + y$ by adding their own two shares locally, and shares of $x \times y$ by executing an interactive protocol that requires exchanging O($\ell$) bits. For boolean secret sharing,
$x \oplus y$ is local, like arithmetic addition, while $x \land y$ requires interaction, like arithmetic multiplication. Given protocols for secure addition and multiplication (in arithmetic sharing) or secure XOR and AND (in boolean sharing), parties can evaluate arbitrary  functions on secret-shared inputs. 

\stitle{Security guarantees.} \ours is designed to provide the \emph{full} security guarantees of the underlying MPC protocols. It protects all input, intermediate, and output data and does not leak any information to untrusted entities, not even the intermediate or output result sizes. \ours guarantees data privacy even in the presence of powerful adversaries that completely control the network and can compromise $t$ out of $n$ computing parties $(t<n)$. Adversaries can be either \textbf{semi-honest} (also known as ``honest but curious''), who passively eavesdrop on the parties' memory contents, access patterns, and network transmissions, or \textbf{malicious} (also known as ``Byzantine''), who can force the parties they control to deviate from the protocol arbitrarily, e.g., by corrupting shares. We further distinguish two classes of MPC protocols: \textbf{honest-majority} protocols, where the adversary controls $t<n/2$ parties, and \textbf{dishonest-majority} protocols, where the adversary controls $t\geq n/2$ parties. In general, malicious-secure protocols for dishonest majorities are the most expensive. \ours provides malicious security \emph{with abort}, meaning that computing parties stop the computation as soon as they detect cheating to protect data privacy.
\camera{We analyze \ours's security in Appendix \ref{apdx:security-analysis}.}

\stitle{Supported protocols.}
\ours supports protocols from the entire spectrum of threat models and can protect data against both semi-honest and malicious parties in honest and dishonest majorities. Specifically, \ours supports the following state-of-the-art protocols: ABY~\cite{aby} (semi-honest, dishonest majority), ABY3~\cite{araki2016high,aby3} (semi-honest, honest-majority), Fantastic Four~\cite{dalskov2021fantastic,cryptoeprint:2026/234} (malicious, honest-majority), and SPDZ2k~\cite{spdz-2k} (malicious, dishonest-majority).
\section{\ours's secure abstractions}\label{sec:abstractions}
We first present \ours's oblivious data abstractions and describe the core interfaces and utilities the system exposes to users who wish to add a new MPC protocol. Then, we discuss \ours's intermediate layer of protocol-agnostic secure functionalities that can be used as black-boxes in higher-level secure operators and analytics libraries.

\subsection{Oblivious data abstractions}\label{sec:vectors}
\ours offers distinct data abstractions for each of its target users (\S\ref{sec:users}), all of which fundamentally rely on \emph{base vector containers}. \ours's data types support a predefined set of operations on their entire content and do \emph{not} allow access to individual elements. This vector-centric design is a core architectural choice in \ours: unlike general-purpose MPC compilers~\cite{keller2020mp,picco} that transform arbitrary programs into oblivious ones (e.g., by multiplexing conditionals or applying  vectorization), \ours operators work on vectors by construction, enabling efficient parallelization, message batching, and memory management under the covers, as we explain in \S\ref{sec:runtime}. 

\stitle{Vector containers.} \ours defines three vector types: 

A \texttt{Vector} is effectively a wrapper around \texttt{std::vector} that provides standard vectorized operations, such as element-wise addition and multiplication.  It is used for local operations on the stored elements and as input to functions generating secret shares. \ours also provides a \texttt{Matrix} type built on \texttt{Vector} that stores data in row-major order.

An \texttt{EVector} is a multi-vector abstraction, i.e., a container that stores a list of \texttt{Vector} objects and supports the same vectorized operations as \texttt{Vector} but applies them to all internal vectors simultaneously. For example, let \texttt{A} and \texttt{B} be two \texttt{EVector} instances, each one with two internal vectors: \texttt{A}$_1$, \texttt{A}$_2$ and \texttt{B}$_1$, \texttt{B}$_2$ respectively. Evaluating \texttt{A} + \texttt{B} performs element-wise addition on each pair of the corresponding inner vectors, i.e., \texttt{A}${}_1$ + \texttt{B}${}_1$ and \texttt{A}$_2$ + \texttt{B}$_2$.
This abstraction is essential for supporting protocols where each party maintains multiple shares per secret. For instance, replicated secret-sharing schemes~\cite{araki2016high,dalskov2021fantastic} require storing multiple shares per secret value, while malicious-secure protocols store message authentication codes (MACs) alongside shares for verifying data integrity~\cite{damgaard2013practical}. An \texttt{EVector}
is the main programming abstraction for protocol developers (cryptographers) to implement low-level MPC primitives, as we explain in \S\ref{sec:protocol-layer}. 

A \texttt{SharedVector} represents an encoded view of a secret vector as seen by an untrusted party and supports secure MPC operations on its contents (secret shares). Different MPC parties have different views of the same secret vector, and these views may vary across protocols. \ours supports both arithmetic and boolean secret sharing (\S\ref{sec:background}) through \texttt{ASharedVector} and \texttt{BSharedVector}, which inherit from the base \texttt{SharedVector} class. This vector type serves as the core data abstraction for library developers to implement intermediate primitives and high-level operators.

Secure operations on \texttt{SharedVector} appear to users as if they were performed on local vectors. For example, given two \texttt{ASharedVector}s \texttt{v1} and \texttt{v2} of the same length, a secure element-wise multiplication is expressed as \texttt{v1 * v2}, where the \texttt{*} operator triggers a secure multiplication functionality that incurs communication between parties. This complexity is handled transparently by the \ours engine.

\stitle{High-level containers.} While cryptographers and library developers operate on vector containers, data analysts work with higher-level abstractions tailored to specific workload types. \ours provides a \texttt{SecureTable} abstraction for secret-shared relational tables, a \texttt{SecureTS} type for secret-shared time series, and a \texttt{SecureMatrix} for secret-shared matrices, e.g., images. Each abstraction exposes a set of secure transformations that can be composed into declarative analytics programs and executed under MPC. For example, \texttt{SecureTable} provides  relational operators such as \texttt{filter}, \texttt{join}, and \texttt{distinct}; \texttt{SecureTS} provides time-based window operators such as \texttt{tumbling\_window}; and \texttt{SecureMatrix} supports ML inference operations, such as \texttt{conv}, \texttt{pooling}, and \texttt{matmul}. 
All these abstractions are built on top of \ours's containers; therefore, converting from one type to another is straightforward through helper functions, as shown in Listing~\ref{lst:multi-workload}.\vspace{1mm}

\noindent \camera{\textbf{A note on expressivity.} \ours's data abstractions consciously restrict the programming model by not allowing (i) for-loops and (ii) if-conditionals on private data at the operator and analytics layer (see Fig.~\ref{fig:stack}). This design facilitates optimization on vectorized operations rather than general programs, while remaining expressive enough to capture the complex analytics workflows users care about. Many plaintext frameworks, like Spark~\cite{zaharia2012resilient}, Flink~\cite{carbone2015apache}, and Google Dataflow~\cite{google-dataflow}, follow a similar declarative approach, where users cannot explicitly iterate over the elements of a data collection and instead apply per-element (e.g., map, filter) or per-group transformations (e.g., reduce, join). \ours's built-in operators allow the same on \texttt{SecureTable} and \texttt{SecureTS}.  Experts can implement custom operators a level below, but these operators must still conform to the higher-level API. Finally,  we note that not allowing conditionals on private data does not mean that \ours cannot express filter predicates or conditional aggregations. Rather, data-dependent branching must be expressed with the provided oblivious operator, which evaluates both branches and assigns results using a secure multiplexer circuit.}

\subsection{MPC protocol abstractions}\label{sec:protocol-layer}
A secure MPC protocol in \ours involves local computation, inter-party communication, and randomness generation.
Adding a new protocol to the system requires defining the share type \texttt{T}, the number of parties $k$, and the small set of protocol-aware primitives shown inside the dark green box of Fig.~\ref{fig:stack} (we provide the full interface in Appendix~\ref{apdx:interface}). 

Functionalities are implemented as methods of an abstract protocol class that encapsulates two key modules: a \emph{communicator} for managing inter-party message exchange, and a \emph{randomness generator} for producing protocol-specific randomness. 
By inheriting from this abstract class, developers gain access to the two modules and can focus exclusively on the protocol logic. Listing \ref{lst:2pc-protocol-multiply} shows an example implementation of the multiplication functionality for a semi-honest 2-party computation protocol (ABY)~\cite{aby}.

Secure multiplication in ABY proceeds in three phases (lines~\ref{lst:mul-start}-\ref{lst:mul-end}). First, each party invokes the randomness generator \texttt{BTGen} to retrieve the required number of Beaver triples (line~\ref{lst:triples}). These are precomputed random triples of the form \texttt{(a, b, c)}, where \texttt{c = a $\times$ b}, and \texttt{a}, \texttt{b}, \texttt{c} are secret-shared among parties using arithmetic sharing, just like inputs \texttt{x} and \texttt{y}. Next, each party randomizes \texttt{x} and \texttt{y} using \texttt{a} and \texttt{b}, respectively, and reveals the masked values to the other party (lines~\ref{lst:masking-start}-\ref{lst:masking-end}). This is done through a user-defined \texttt{reconstruct()} function, which internally invokes the communicator. Once the exchange completes, parties perform a final local operation that computes and unmasks the shares of the result (line~\ref{lst:unmask}). 
Although the entire code appears to operate on individual shares, all steps are in fact applied to vectors using \texttt{EVector}'s overloaded operators~($+, -, *$). 

\begin{lstlisting}[language=C++, caption={Implementation of the ABY~\cite{aby} secure multiplication functionality in \ours. Each party  executes the function on its own arithmetic shares of \texttt{x} and \texttt{y} to end up with shares of the result~\texttt{z}.}, numbers=left, label={lst:2pc-protocol-multiply}, escapechar=|]
// User-defined helper function
Vector reconstruct(const EVector& shares) {|\label{lst:open-start}|
	Vector remote_shares(shares(0).size());//Receive buffer|\label{lst:acc1}|
	// Exchange shares with other party
	this->communicator->exchange_shares(shares(0), |\label{lst:acc2}|
																			remote_shares, +1);
	return shares(0) + remote_shares;	// Reconstruct
}|\label{lst:open-end}|

// Multiplies x, y elementwise and stores result in z
void multiply_a(const EVector& x, const EVector& y,    
								EVector& z) {|\label{lst:mul-start}|
	// Step 1: Pull arithmetic Beaver triples (c = a * b)
	auto [a, b, c] = this->BTgen->getNext(x.size());|\label{lst:triples}|
	// Step 2: Reveal randomized inputs (interactive)
	auto r|$_\texttt{x}$| = reconstruct(x + a);|\label{lst:masking-start}|
	auto r|$_\texttt{y}$| = reconstruct(y + b);|\label{lst:masking-end}|
	// Step 3: Compute multiplication result (local)
	z = y*r|$_\texttt{x}$| - a*r|$_\texttt{y}$| + c; // z = x * y (noise cancels out)|\label{lst:unmask}|
}|\label{lst:mul-end}|
\end{lstlisting}

The \texttt{()} operator in lines~\ref{lst:acc1}~and~\ref{lst:acc2} is used to access the \texttt{EVector}'s contents; that is, \texttt{shares(0)} returns the first inner vector of the \texttt{shares} container (at index 0).
Recall from \S\ref{sec:vectors} that \texttt{EVector} is a multi-vector abstraction containing one or more inner \texttt{Vector}s whose number is protocol-specific. In ABY, each party receives one share of each input secret and, hence, the \texttt{EVector} type contains a single inner vector. \texttt{EVector} dimensions are set by protocol developers as a template parameter when instantiating the protocol class. 

\stitle{Supporting asymmetric protocols.}
ABY is a symmetric protocol, meaning that both parties execute the same code on different input shares. Other protocols, however, are inherently asymmetric and require parties to perform different operations based on their roles in the computation. To support such protocols, \ours assigns each party a unique \texttt{partyID} in $[0, k)$, where $k$ is the number of parties, that developers can access through the protocol class to implement role-specific logic. For example, a developer can write \texttt{if(partyID==1)~\{...\}} to condition execution on a specific party. We use this technique to implement the asymmetric Fantastic Four protocol~\cite{dalskov2021fantastic}.

\stitle{Expressing communication patterns.}
In line~\ref{lst:acc2} of Listing~\ref{lst:2pc-protocol-multiply}, parties exchange messages by invoking the communicator's \texttt{exchange\_shares()} method with three arguments: a \texttt{Vector} of shares to send, an empty \texttt{Vector} to store the received shares, and a signed integer specifying the destination party. \ours organizes parties in a \emph{logical ring} topology, allowing developers to refer to other parties using relative indices: the notation \texttt{+i} denotes the $i$-th successor on the ring. In our 2-party example, \texttt{+1} indicates that each party sends to its successor and receives from its predecessor.

To support more complex communication patterns, \ours provides a generalized \texttt{exchange\_shares()} method that accepts four parameters: a collection of vectors to send, a collection of vectors to populate with received data, the relative indices of destination parties, and the relative indices of source parties. Protocol developers can use this method along with the \texttt{partyID} to implement arbitrary communication patterns. For example, assuming three parties on the ring, a multicast operation where party 0 sends the same vector \texttt{S} to parties 1 and 2 is implemented as shown below: 

\begin{lstlisting}[language=C++, caption={}, numbers=left, label={lst:multicast}, escapechar=|]
	if (partyID==0) { // Send vector S to parties 1 and 2
		exchange_shares({S, S}, {}, {+1,+2}, {});
	}
	if (partyID==1) { // Receive into vector R from party 0
		exchange_shares({}, {R}, {}, {-1}); // +2 also works
	}
	if (partyID==2) { // Receive into vector R from party 0
		exchange_shares({}, {R}, {}, {-2});	// +1 also works
	}
\end{lstlisting}

Note that each call to \texttt{exchange\_shares()} is blocking to ensure that interacting parties receive each other's shares before proceeding. However, send and receive operations are handled by dedicated communication threads (see \S\ref{sec:runtime}) and are performed in parallel. 

\stitle{Randomness generation.}
Many MPC protocols rely on a data-independent offline phase known as preprocessing, which generates randomness that is then used to speed up the online phase. The arithmetic triples used in Listing~\ref{lst:2pc-protocol-multiply} are an example of correlated randomness.
\ours provides access to pseudorandom number generators (PRGs) based on AES, zero-sharing generators needed by some honest-majority protocols~\cite{araki2016high,dalskov2021fantastic}, random permutations needed for 2-party oblivious shuffling~\cite{permutation-correlations}, and Beaver triples~\cite{Beaver95,Beaver91a} (both boolean and arithmetic). 
Triples can be generated on-the-fly (expensive) or precomputed and stored offline, and then consumed during the online phase as needed. 

\stitle{Supporting other protocol types.}
\ours's \texttt{EVector} abstraction is designed for protocols where each input secret is encoded as one or more shares; however, some protocols deviate from this paradigm. A representative example is MiniMac~\cite{damgaard2013constant}, which encodes multiple input secrets into a single (array-like) shared representation. Adding such a protocol to \ours requires developers to implement two additional functionalities: (i) an extraction function that defines how to retrieve a single element encoding from a packed representation, and (ii) a bit extraction function that defines how to extract individual bits needed for boolean operations.

\subsection{Protocol-agnostic primitives}\label{sec:intermediate-primitives}
\ours provides a set of protocol-agnostic secure primitives that use the underlying MPC functionalities as black boxes. The intermediate primitives include secure multiplexers, ripple-carry and parallel-prefix adders, comparison operators $(>, <, \geq, \leq, ==, \neq)$, compare-and-swap, ReLU, convolution, average pooling, dot product, and matrix multiplication. 
These primitives serve as building blocks in higher-level operators across all analytics libraries in \ours. Developers can selectively override the default implementations with more efficient protocol-specific variants.
For example, the default dot product operation is evaluated by invoking element-wise secure multiplication followed by aggregation. However, in replicated secret sharing protocols~\cite{araki2016high,dalskov2021fantastic}, we can first execute the local part of the secure multiplication for all elements, then aggregate locally, and perform the interactive part of the multiplication in the end. This approach reduces communication complexity from linear to constant.
\begin{figure}[t]
  \centering
      \includegraphics[width=\linewidth]{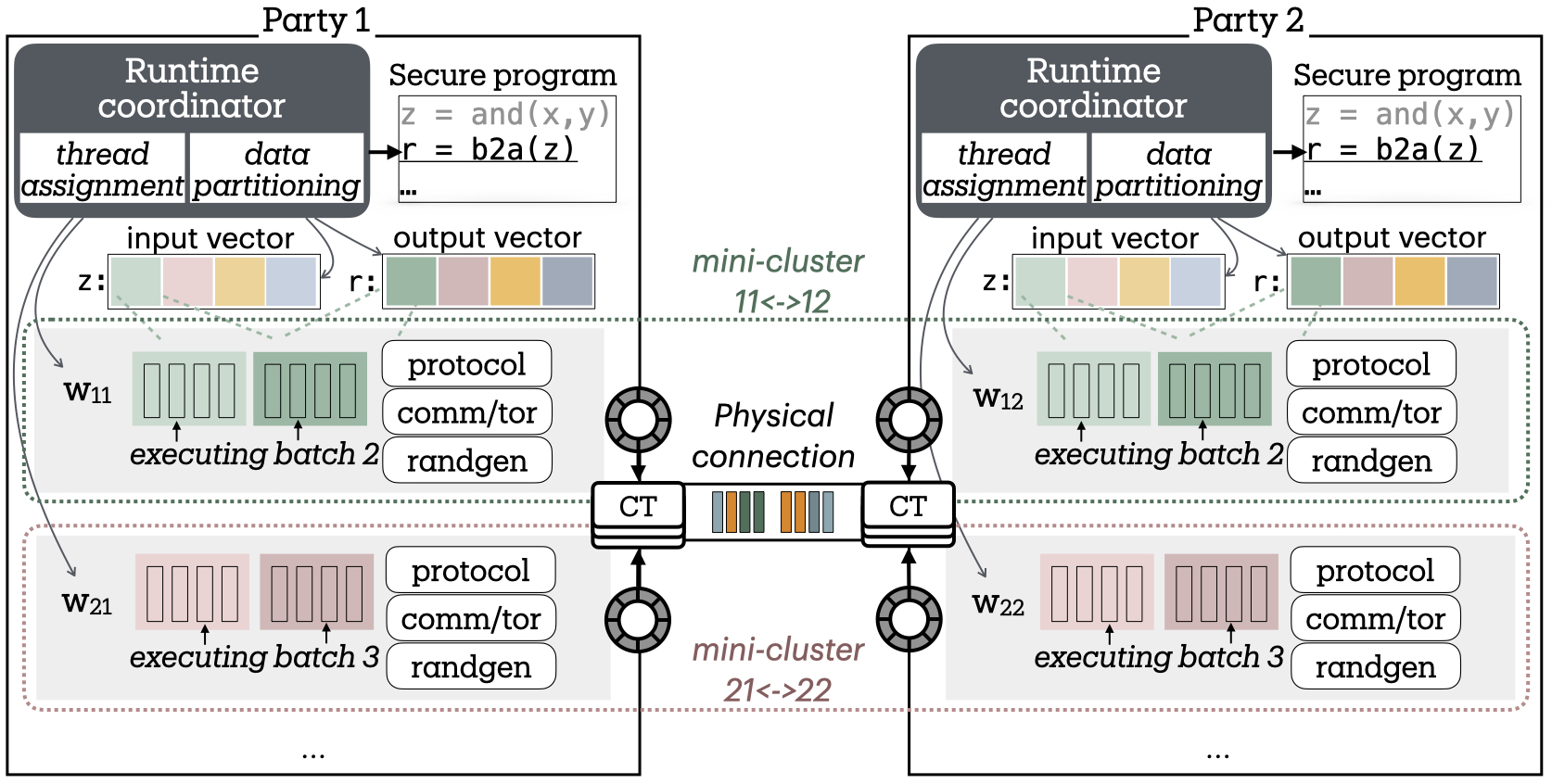}
  \caption{The \ours execution model shown for 2 parties. Each party runs a \ours process with a runtime coordinator that handles data partitioning and work assignment. Pairs of worker threads ($w_{ij}$) across parties form independent computation "mini-clusters".  Message exchange is handled by communication threads (CTs) \camera{and multiplexed over physical pairwise connections.}}
  \label{fig:runtime}
\end{figure}

\section{Data parallelism and virtual vectors}\label{sec:runtime}

The vectorized abstractions described above, which require all computations to be defined as element-wise operations on vectors, provide two major benefits: (i) they allow message batching across operations to amortize MPC communication costs, and (ii) they enable efficient workload distribution by partitioning vectors across parallel workers. In this section, we first describe  \ours's execution model and then explain how constructing \emph{virtual vectors} can transparently extend parallelization to high-level user programs.

\subsection{Execution model}\label{sec:execution-model}
\ours's execution model is shown in Fig.~\ref{fig:runtime} for an example 2-party MPC protocol.
Each \ours party runs on a physical machine (or VM or container) connected to other parties through a network. \ours represents each party as a single OS process that includes a coordinator thread (the runtime) and a set of worker threads.  Each worker thread on one party forms a \emph{mini-cluster} with its corresponding partner threads on the other parties.  Mini-clusters operate independently, maintaining separate protocol, communicator, and randomness generator objects to execute secure computations. Communication between threads within a mini-cluster is multiplexed over the  physical network connection through logical channels. To avoid data copies, dedicated communication threads (CTs) manage message passing using lock-free buffers shared with mini-cluster computation threads.

All parties execute the same program as a sequence of secure primitives that operate on vectors.  For each primitive, the runtime coordinator handles memory allocation, partitions the input and output vectors into disjoint subsets, and assigns each partition to a mini-cluster. Each mini-cluster further divides its partition into smaller batches.  Batches within a mini-cluster are processed in order, and mini-clusters synchronize only at primitive boundaries: all mini-clusters must complete the current primitive before  proceeding to the next (barrier). However, different mini-clusters may process different batches within the same primitive concurrently. This design overlaps local computation with communication across mini-clusters, improving resource utilization.

It is important to note that a secure primitive (all operations in the protocol layer of Fig.~\ref{fig:stack}) represents the unit of computation assigned to a mini-cluster. A thread executes the entire control flow of a primitive on its assigned partition, potentially involving multiple communication rounds with partner threads running in remote machines, but without coordination with local worker threads. This way, high-level operators, such as sorting or aggregation, can also be transparently parallelized by the runtime, as long as they are expressed as compositions of secure vectorized primitives. We describe how \ours achieves this next. 

\subsection{The need for vectorization}
Let us first illustrate why \ours enforces vectorized primitives and their composition by design. Consider the oblivious compare-and-swap operation shown in Listing~\ref{lst:for-loop}, implemented in an imperative style, where secure operations, such as comparison ($<=$),  XOR (\textasciicircum), and AND ($\&$) are applied to individual elements.  Function \texttt{extend\_lsb()} constructs a mask of type \texttt{T} whose $\ell$ bits equal the single-bit result of the respective comparison.  Note that this code snippet does not correspond to valid \ours code, as the framework does not allow iterating over individual elements of \texttt{BSharedVector}~(\S\ref{sec:abstractions}). 

The problem with this implementation is that some secure operations, such as comparison and AND, require communication between parties. Performing these operations sequentially results in a large number communication rounds, where each round acts as an expensive  synchronization barrier in the distributed execution.

\begin{lstlisting}[language=C++, caption={Baseline compare-and-swap primitive that uses non-vectorized secure operators (in orange). Each comparison (\textcolor{orange}{\texttt{<=}}) requires O($\log|\texttt{T}|$) rounds, where |\texttt{T}| is the length of the secret share representation in bits. For $n$-sized input vectors, the primitive performs O($n\log|\texttt{T}|$) communication rounds in total.}, numbers=left, label={lst:for-loop}, escapechar=|]
void non-vectorized-cmp-swp(BSharedVector<T>& v1, 
													  BSharedVector<T>& v2){
	int n = v1.size();		// v1 and v2 have equal length
	BSharedVector mask(n);
	// Apply elementwise comparisons and compute masks 
	for (int i=0; i<n; i++) { // O(n|\textcolor{codegreen}{$\log\vert\texttt{T}\vert$}|) comm. rounds
		mask[i] = (v1[i] |\textcolor{orange}{\textbf{<=}}| v2[i]).extend_lsb(); 
	}	
	// Swap i-th elements obliviously if v1[i] <= v2[i]
	BSharedVector tmp(n);
	for (int i=0; i<n; i++) { // O(n) comm. rounds
		tmp[i] = mask[i] |\textcolor{orange}{\textbf{\&}}| (v1[i] |\textcolor{orange}{\raisebox{0.5ex}{\scalebox{0.7}{$\mathbf{\wedge}$}}}| v2[i]);
		v1[i] |\textcolor{orange}{\raisebox{0.5ex}{\scalebox{0.7}{$\mathbf{\wedge}$}}\textbf{=}}| tmp[i];
		v2[i] |\textcolor{orange}{\raisebox{0.5ex}{\scalebox{0.7}{$\mathbf{\wedge}$}}\textbf{=}}| tmp[i];
	}
}
\end{lstlisting}

Although some MPC compilers can recognize such patterns \cite{keller2020mp}, it is difficult to guarantee that compiler transformations will optimize arbitrary user code. 
In contrast, exposing a vectorized API enables \ours to execute the local computation for multiple element-wise operations in parallel and exchange the resulting messages within a single network call.
Listing~\ref{lst:cmp-swp} shows the same compare-and-swap program written using \ours's vectorized primitives~(in blue).  

\begin{lstlisting}[language=C++, caption={\ours's compare-and-swap primitive. Vectorized secure operators (in blue) call into the system runtime that manages message batching under the hood. This implementation consumes the same bandwidth as Listing~\ref{lst:for-loop}, but performs O($\log|\texttt{T}|$) communication rounds in total, independent of the input~size.}, numbers=left, label={lst:cmp-swp}, escapechar=|]
void |\textbf{cmp-swp}|(BSharedVector<T>& v1, BSharedVector<T>& v2){
	// Compare and compute mask - O(|\textcolor{codegreen}{$\log\vert\texttt{T}\vert$}|) comm. rounds
	BSharedVector<T> mask = (v1 |\textcolor{blue}{\textbf{\texttt{<=}}}| v2).extend_lsb();
	// Swap i-th elements obliviously if v1[i] <= v2[i]
	auto tmp = mask |\textcolor{blue}{\textbf{\texttt{\&}}}| (v1 |\textcolor{blue}{\raisebox{0.5ex}{\scalebox{0.7}{$\mathbf{\wedge}$}}}| v2); // O(1) comm. rounds
	v1 |\textcolor{blue}{\raisebox{0.5ex}{\scalebox{0.7}{$\mathbf{\wedge}$}}\textbf{=}}| tmp;  // v1[i] = v1[i] ^ tmp[i]
	v2 |\textcolor{blue}{\raisebox{0.5ex}{\scalebox{0.7}{$\mathbf{\wedge}$}}\textbf{=}}| tmp;  // v2[i] = v2[i] ^ tmp[i]
}
\end{lstlisting}

\begin{figure}[t]
\centering
\begin{subfigure}[b]{0.44\textwidth}
    \centering
    \includegraphics[width=\textwidth]{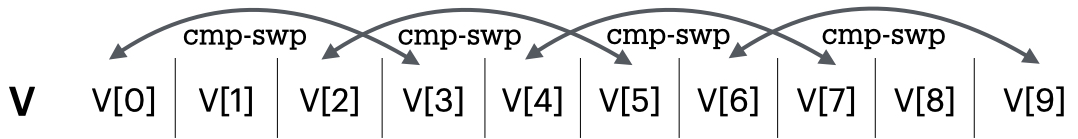}
    \caption{An example oblivious control flow that compares every other pair of elements at distance $d=3$ in vector $V$. Splitting $V$ into disjoint parts results in comparisons spanning thread-local boundaries. }
    \label{fig:vv-control-flow}
\end{subfigure}
\hfill
\begin{subfigure}[b]{0.44\textwidth}
    \centering
    \includegraphics[width=\textwidth]{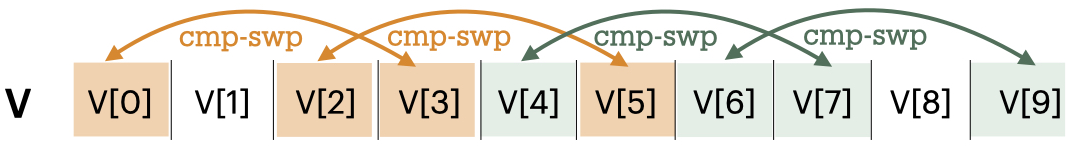}
    \caption{The strawman approach to apply \ours's vectorized \texttt{cmp-swp()} would require writing custom logic to extract elements into temporary vectors, split them across workers, and copy results back into $V$.}
    \label{fig:vv-threads}
\end{subfigure}
\hfill
\begin{subfigure}[b]{0.44\textwidth}
    \centering
    \includegraphics[width=\textwidth]{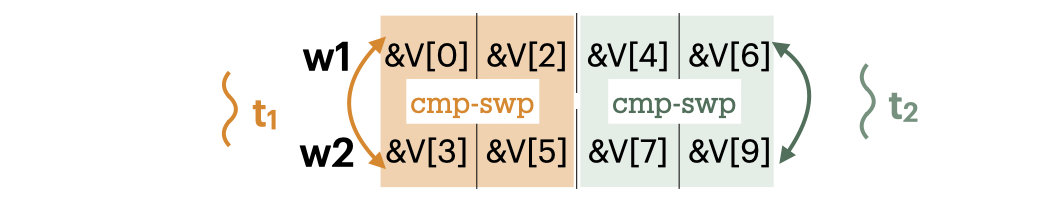}
    \caption{Virtual vectors $w_1$, $w_2$ are partitioned by the runtime automatically.}
    \label{fig:vv-result}
\end{subfigure}
\caption{Virtual vectors allow transparent reuse of vectorized primitives in high-level operators and automatic parallelization.}
\label{fig:vv-example}
\end{figure}

\subsection{Virtual vectors}\label{sec:virtual-vectors}

While the need for vectorized operations to enable batching has been recognized by prior systems~\cite{liagouris2023secrecy}, the key challenge we address in \ours lies in enabling \emph{transparent reuse} of these primitives within higher-level operators. Consider the example oblivious control flow shown in Fig.~\ref{fig:vv-control-flow}, which compares and swaps every other pair of elements at distance $d=3$ in a vector $V$. Such data access patterns appear frequently in oblivious prefix~\cite{hillis1986data}, sorting~\cite{Batcher68}, and butterfly-style networks~\cite{B_goodrich2011data} that form the basis of many high-level data operators.  
This type of operators do not conform to a regular element-wise access pattern but instead require accessing non-contiguous elements of the same vector.

Implementing the example of Fig.~\ref{fig:vv-control-flow} using \ours's vectorized API would require traversing the input vector to extract the respective elements, copying them to a new vector, and then applying the vectorized compare-and-swap primitive.  Parallelizing the computation is another challenge, as simply partitioning the vector does not work. Instead, a developer would need to implement the task-to-worker assignment logic shown in Fig.~\ref{fig:vv-threads} (see Appendix~\ref{apdx:custom-example} for an example implementation).  Overall, developers would need to write custom data transformation and parallelization code for every high-level operator, leading to inefficient and error-prone implementations that hinder codebase maintenance.

\begin{lstlisting}[language=C++, caption={Data-parallel implementation of the oblivious operator from Fig.~\ref{fig:vv-example} in \ours. Virtual vectors enable library developers to write single-threaded code while pushing all complexity of parallelization and memory management to the system runtime.}, numbers=left, label={lst:vv-example}, escapechar=|]
// Both virtual vectors w1 and w2 point to elements in V
BSharedVector<T> w1 = V.limit(7).step(2); // Left inputs
BSharedVector<T> w2 = V.offset(3).step(2);// Right inputs
|\textbf{cmp-swp}|(w1, w2); // Updates V in place (no extra copies)
\end{lstlisting}

To address this challenge, \ours provides a set of methods that allow developers to define \emph{virtual vectors}. A virtual vector $W$ is a restricted view of a base vector $V$ such that $W[i] = V[f(i)]$, where $f$ defines a mapping from indices in $W$ to indices in $V$. A base vector can have multiple virtual vectors associated with it, all pointing to elements of the same underlying array (``shallow copies''). Virtual vectors are constructed via a set of functions that allow declarative expression of irregular access patterns on other vectors. 

Listing~\ref{lst:vv-example} and  Fig.~\ref{fig:vv-result} show the same oblivious control flow expressed using virtual vectors.
Functions \texttt{offset()}, \texttt{step()}, and \texttt{limit()} define virtual vectors on $V$: \camera{$w1$} that steps every 2 elements up to index 7, and \camera{$w2$} that starts at $V[3]$ and steps every 2 elements up to $V$'s size (limits are exclusive). 
These two vectors are passed as input to the \texttt{cmp-swp} primitive, which performs the required element-wise comparisons and updates the base vector $V$ in place. 

The \ours runtime partitions and distributes virtual vectors as if they were standard (contiguous) vectors, automatically parallelizing user code while handling index translations and memory management without data copies. Thanks to the deterministic nature of oblivious programs, index translation can be computed analytically. Virtual vectors can be defined on other virtual vectors to express more complex control flows, like cyclically traversing every other element of a base vector. In Appendix~\ref{apdx:access-patterns}, we summarize the core data access patterns and corresponding virtual vectors that are sufficient to express all oblivious control flows in \ours. We also provide their mapping functions, interfaces, and an example.

\subsection{Discussion}
Virtual vectors share conceptual similarities with hierarchical IRs for GPU programming~\cite{cuda-tile, graphene}, which enable efficient thread binding without data copies or expensive synchronization.  \ours virtual vectors are a flat data representation tailored to the characteristics of oblivious programs, where deterministic control flow enables index computation at compile time and automatic parallelization at runtime.

This design relies on a key property of oblivious algorithms: \emph{conflict-free data access}. 
One might question how \ours handles potential conflicts, but in practice, oblivious control flows are structured into well-defined stages where computation within each stage is embarrassingly parallel by design. Derived from HPC and hardware literature~\cite{KUNG198065}, these algorithms avoid expensive synchronization within each stage, and when index overlaps occur (e.g., in odd-even merge~\cite{DBLP:journals/iacr/JonssonKU11}),
the conflicting operations are read-only. Providing compile-time or runtime verification that developers use virtual vectors correctly is an interesting direction for future work; \ours does not currently enforce such constraints.

As a final comment, virtual vectors enable automatic parallelization for a broad class of oblivious operators; however, there exist cases where we cannot easily leverage them.  An example is oblivious shuffling~\cite{ahi22-radixsort}, which requires random data accesses in the underlying vector. \ours can still parallelize parts of the shuffling protocol using virtual vectors, but most of the work in this case runs single-threaded.
\section{System implementation}\label{sec:impl}
\ours is the result of a multi-year, collaborative development effort involving systems researchers and cryptographers. The codebase is written in C++ 
and uses \texttt{libsodium} \cite{libsodium} for pseudorandom number generation, \texttt{libOTE}~\cite{libote} for MPC pre-processing, and Blaze~\cite{blaze-lib} for local matrix operations. The system provides a generic, extensible communicator interface, currently supporting MPI and a socket-based implementation.  \ours adopts a polymorphic design that uses C++ templates to enable custom share representations and compile-time MPC protocol selection. 

\stitle{Implementation details.} Our current implementation of the virtual vector mechanism (\S\ref{sec:virtual-vectors}) precomputes and materializes index mappings to remove index translation from the critical path of the computation. We are also investigating alternative approaches, including dynamic switching between iterator-based, index precomputation, \camera{and view materialization} techniques, as well as expression templates~\cite{iglberger2012high, iglberger2012expression} for efficient operator composition.
\ours also employs bit compression for boolean operations, which typically operate on a few bits per secret-shared element. The system provides functions to extract bits into compressed vectors and  reinsert them into their original positions.  

\stitle{Practical considerations.} \ours can be deployed over LAN or WAN, either on bare-metal machines or cloud resources. Due to the communication-intensive nature of MPC, tuning parameters such as batching and the number of network threads can significantly impact performance.  Our experience suggests that more batches per mini-cluster (\S\ref{sec:execution-model}) benefit LAN deployments by enabling computation-communication overlap, while WAN deployments benefit from fewer batches and more communication threads (CTs) to minimize expensive round trips under high RTT. By default, \ours uses 12 batches with 4 CTs in LAN, and 1 batch with 16 CTs in WAN.  

\stitle{Scalability and fault tolerance.} \camera{\ours uses a single server per party and restarts the computation in case of failures. We note that real MPC deployments~\cite{mpc-use-cases} involve very few parties (hence, the failure rate is low), and this is often mandated by the high communication cost of MPC. All state-of-the-art MPC systems focus on the outsourced setting that uses a small number of untrusted servers (in practice, machines managed by different infrastructure providers) to support any number of data owners, who outsource secret shares of their data to the servers. This is the default setting in \ours and all baselines used in the paper.  So far, horizontal scaling and failures have not been a main concern in the literature, since research on MPC systems is still in early stages.  Adding support for fault-tolerance and distributed execution within each logical party are exciting directions for future work. } 

\section{Evaluation}\label{sec:evaluation}

Our experimental evaluation consists of three parts:

\begin{itemize}
\item In \S\ref{sec:exp-mixed}, we evaluate \ours on a real-world pipeline that includes ML, relational, and time series operations. Our results show that \ours provides practical performance for multi-workload analytics across the threat model spectrum.

\item In \S\ref{sec:exp-comp}, we compare \ours with five state-of-the-art MPC systems. We show that \ours is competitive with specialized analytics frameworks and outperforms the popular MP-SPDZ compiler by up to $4.7\times$. 

\item In \S\ref{sec:exp-scalability}, we evaluate secure primitives and high-level operators in isolation and show that \ours can effectively parallelize computations at all layers of the stack.

\end{itemize}

\stitle{Experimental setting.} 
We run experiments on four environments: (i) \texttt{bm-LAN} and (ii) \texttt{bm-WAN} use a bare-metal cluster consisting of 4 AMD EPYC 9655P 96-core machines; (iii) \texttt{AWS-LAN} and (iv) \texttt{AWS-WAN} use two \texttt{g6.8xlarge} AWS instances with NVIDIA L4 GPUs. LAN environments have up to $25$Gbps bandwidth and $0.17$-$0.25$ms RTT, while in the corresponding WAN environments we restrict the network characteristics to $6$Gbps bandwidth and $20$ms RTT. All nodes run Linux Ubuntu 24.04.4 LTS. Unless otherwise specified, we use $64$-bit shares for \ours and all competitors. Reported numbers are the average of three runs. For dishonest-majority protocols, we only report online times.

\begin{figure}[t]
    \centering
    \begin{minipage}[b]{0.42\textwidth}
        \centering
        \includegraphics[width=\textwidth]{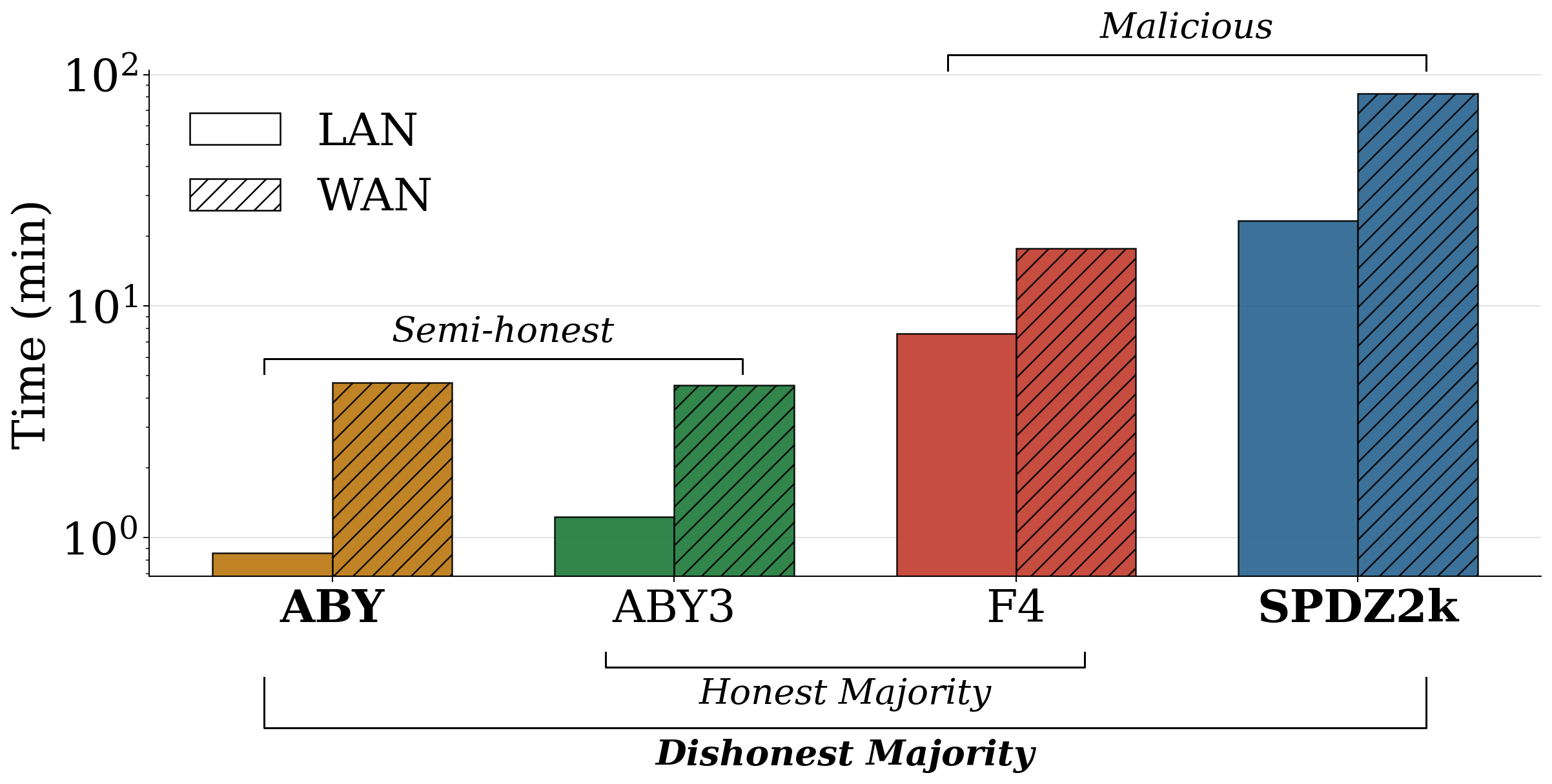}
    \end{minipage}\vspace{-4mm}
    \caption{\ours evaluation on the complex analytics pipeline of Fig.~\ref{fig:mixed-pipeline} across all threat models and network settings. }\label{fig:multi-workload}
\end{figure}

\subsection{Performance on mixed workloads}\label{sec:exp-mixed}
Our first experiment evaluates \ours's capability to support end-to-end secure, mixed-analytics workflows across the entire spectrum of MPC threat models. We implement the motivating use case of Fig.~\ref{fig:mixed-pipeline}, which consists of four stages: (i) image classification on a Kaggle dataset with 64 X-Ray images using a fine-tuned VGG16 model architecture~\cite{xray-vgg16}, (ii) a relational join of the classification results with $1K$ health records of asthmatic patients, (iii) a threshold window followed by a filter and aggregation on $16K$ $SPO_2$ time series measurements, and (iv) a relational subquery consisting of a join, three filters, and a group-by with aggregation to compute the number of predicted diagnoses per asthma group. Fig.~\ref{fig:multi-workload} shows the end-to-end execution time of the pipeline under all MPC protocols in \ours, in both \texttt{bm-LAN} and \texttt{bm-WAN}.

When configured with semi-honest security, \ours evaluates the entire pipeline in $\approx 1min$ in LAN and under $5min$ in WAN. Malicious security is more expensive but still practical for batch analytics: \ours completes execution in $18min$ under honest majority and just over $1h$ under dishonest majority in WAN. To our knowledge, this is the first demonstration of such a complex, mixed-workload analytics pipeline under MPC, and the first evaluation of secure inference using a large CNN model like VGG16 with~SPDZ2k.

Although we did not attempt to implement a baseline that ``stitches together'' specialized systems (which may be feasible for systems with a common threat model), we can estimate the expected costs of such a solution. There would be three primary sources of overhead: (i) \emph{initialization} time, including spawning new processes, allocating memory, and establishing network connections between parties, (ii) \emph{secret re-sharing} time, required to convert output shares from one system into input shares compatible with the next system's protocol, and (iii) \emph{intermediate result materialization}, where shares must be written to persistent storage (e.g., files) by one system and subsequently loaded by the next.

\ours eliminates these overheads by executing the entire workflow within a unified environment. While the next sections demonstrate that \ours is competitive with state-of-the-art systems on individual workloads, the end-to-end benefits would be substantially more pronounced for mixed pipelines. 

\subsection{Comparison with state-of-the-art}\label{sec:exp-comp}

\begin{table}[t]
\centering\small
\begin{tabular}{c|c|c}
{\bf System }& {\bf MPC Protocols} & {\bf Workload} \\ 
\hline\hline
ORQ~\cite{baum2025orq}    & ABY~\cite{aby}, ABY3~\cite{araki2016high, aby3} & Relational \\ 
TVA~\cite{faisal2023tva}    & ABY3, F4~\cite{dalskov2021fantastic} & Time series \\ 
Pigeon~\cite{harth2025pigeon}  & Trio~\cite{harth2025high} & CNN inference \\ 
Piranha~\cite{watson2022piranha}  & SecureML~\cite{mohassel2017secureml} & CNN inference \\
\hline\hline
MP-SPDZ~\cite{keller2020mp} & SPDZ2k~\cite{spdz-2k} & CNN inference \\
\end{tabular}
\caption{State-of-the art systems along with the MPC protocols and  workloads we use for comparison. MP-SPDZ is a general-purpose MPC compiler with built-in support for CNN inference.}\label{tab:competitors}
\end{table}

\begin{figure*}[t]
    \centering
    \begin{minipage}[b]{0.24\textwidth}
        \centering
        \includegraphics[width=\textwidth]{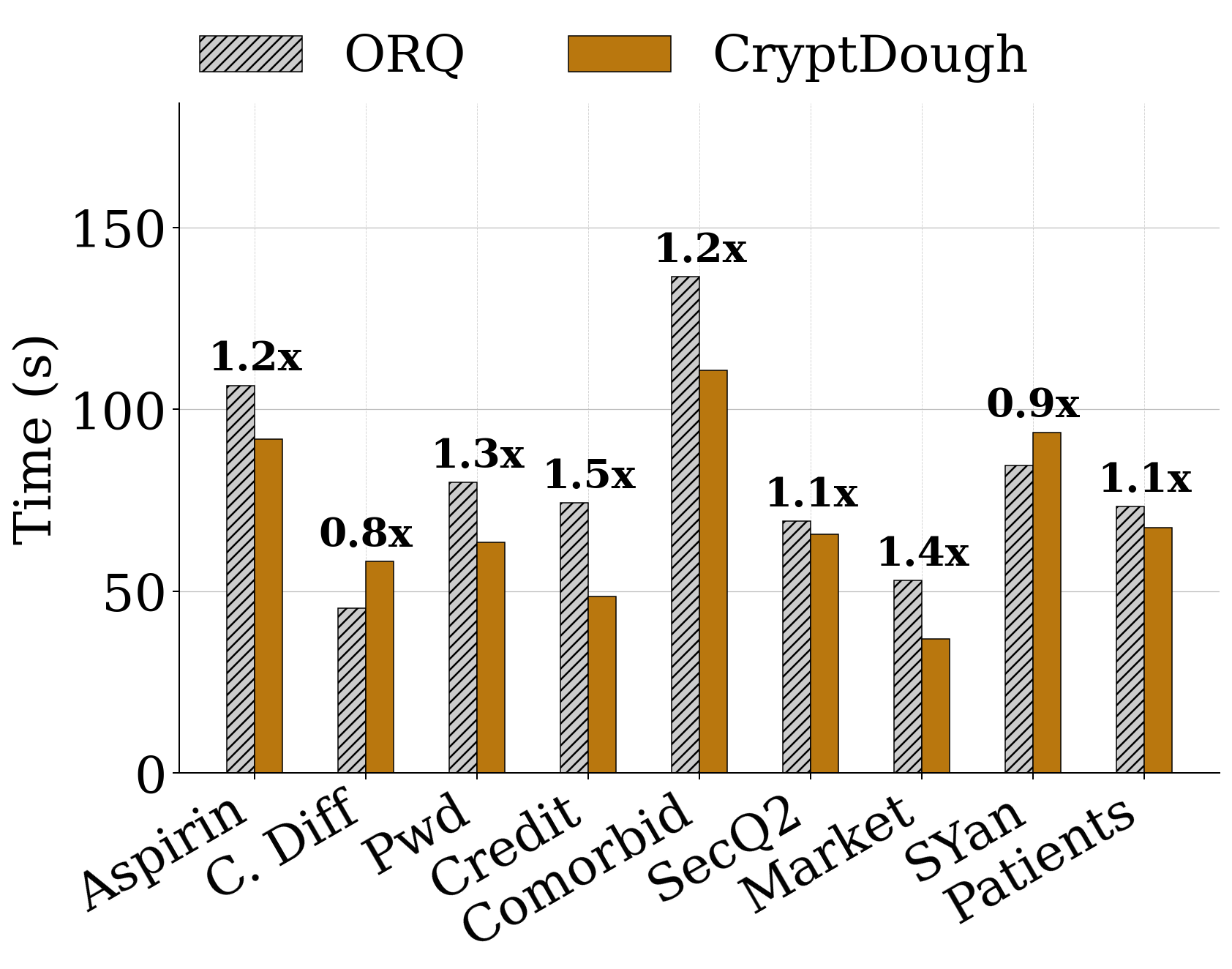}\vspace{-2mm}
        \subcaption{ABY (\texttt{bm-LAN})}
        \label{fig:orq-2pc-lan}
    \end{minipage}
    \hfill
    \begin{minipage}[b]{0.24\textwidth}
      \centering
      \includegraphics[width=\textwidth]{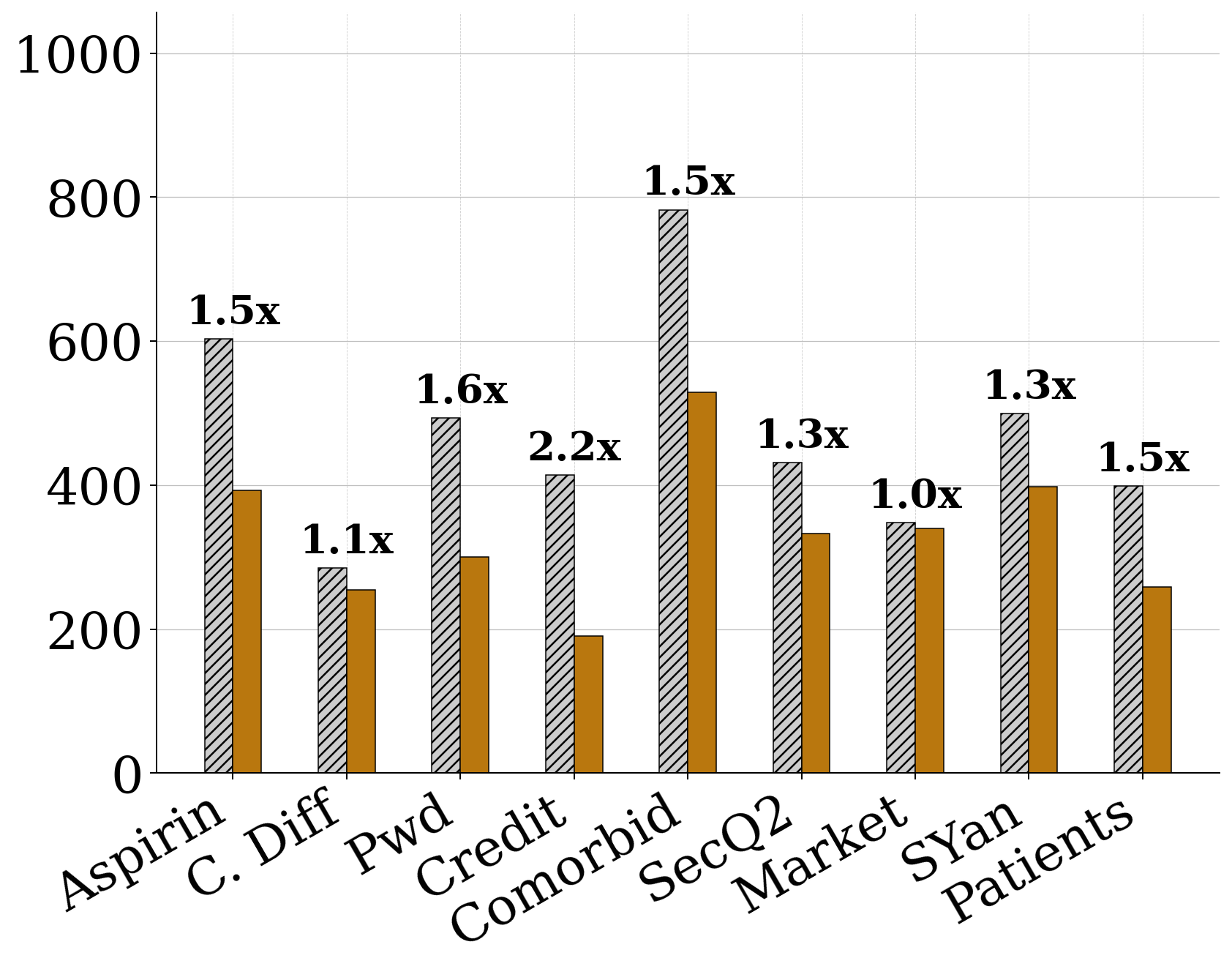}\vspace{-2mm}
      \subcaption{ABY (\texttt{bm-WAN})}
      \label{fig:orq-2pc-wan}
    \end{minipage}
    \hfill
    \begin{minipage}[b]{0.24\textwidth}
        \centering
        \includegraphics[width=\textwidth]{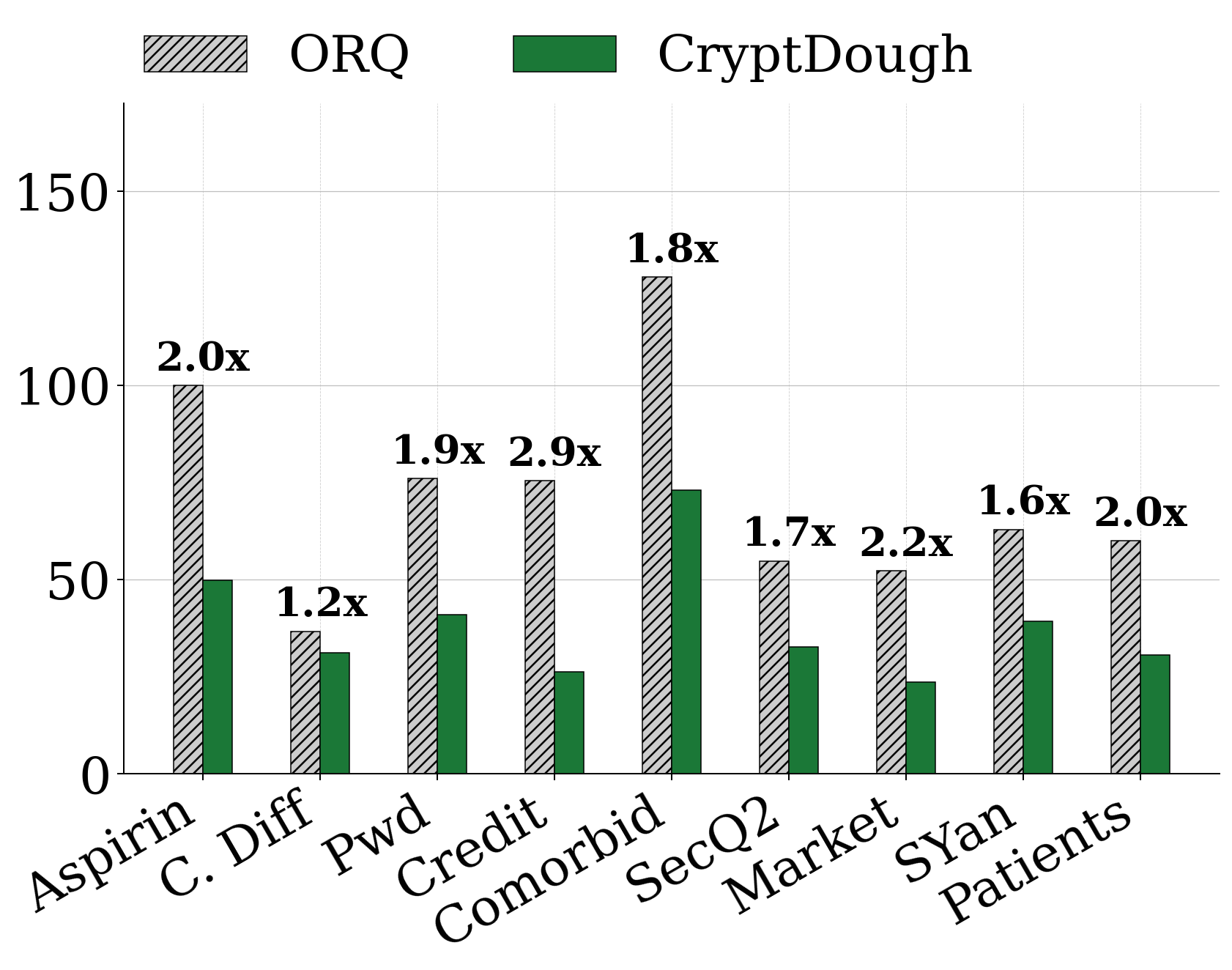}\vspace{-2mm}
        \subcaption{ABY3 (\texttt{bm-LAN})}
        \label{fig:orq-3pc-lan}
    \end{minipage}
    \hfill
    \begin{minipage}[b]{0.24\textwidth}
        \centering
        \includegraphics[width=\textwidth]{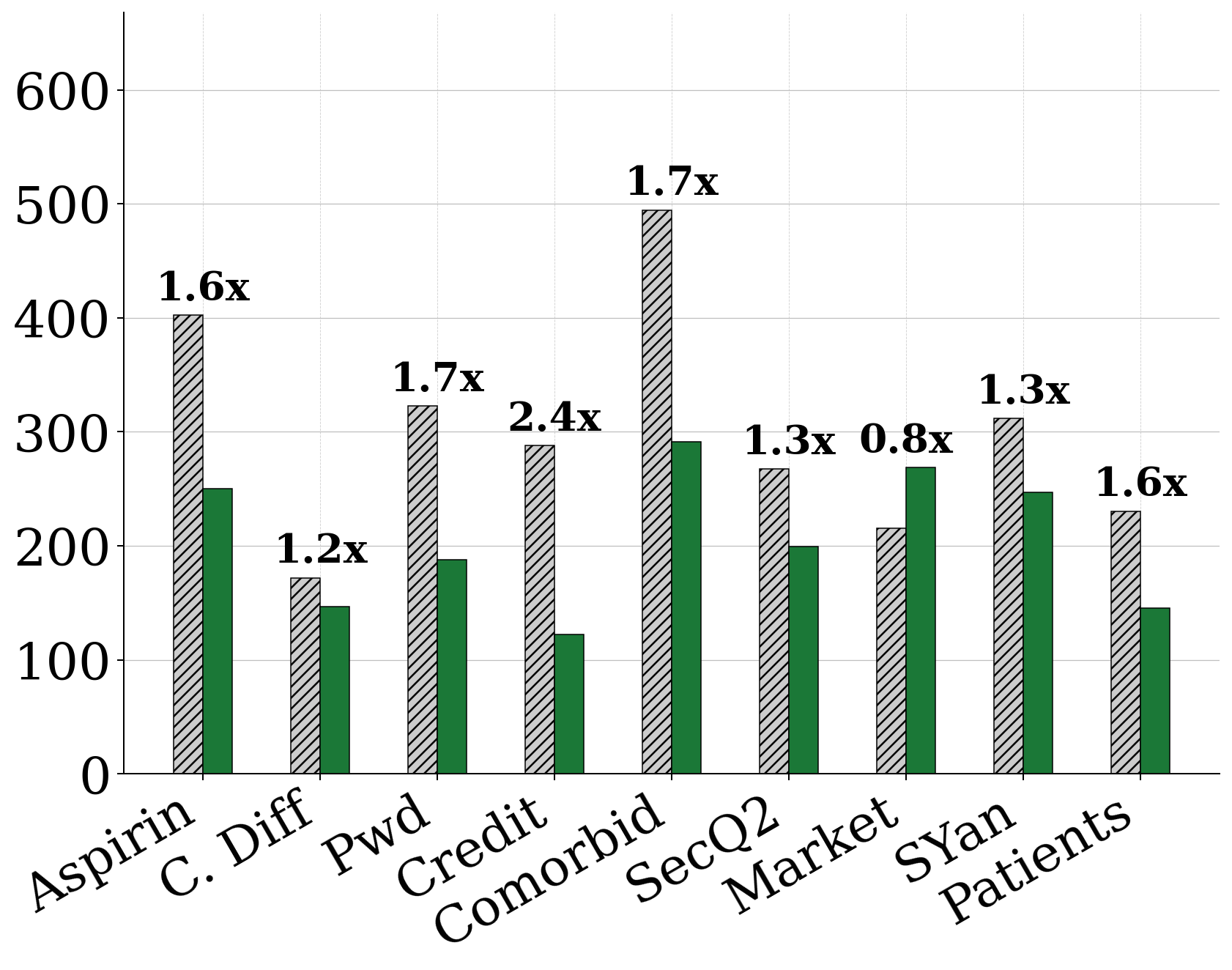}\vspace{-2mm}
        \subcaption{ABY3 (\texttt{bm-WAN})}
        \label{fig:orq-3pc-wan}
    \end{minipage}\vspace{-4mm}
    \caption{\camera{Comparison with ORQ on all real-world queries used in the ORQ paper}}\label{fig:comparison-orq}
\end{figure*}

\begin{figure*}[t]
    \centering
    \begin{minipage}[b]{0.24\textwidth}
        \centering
        \includegraphics[width=\textwidth]{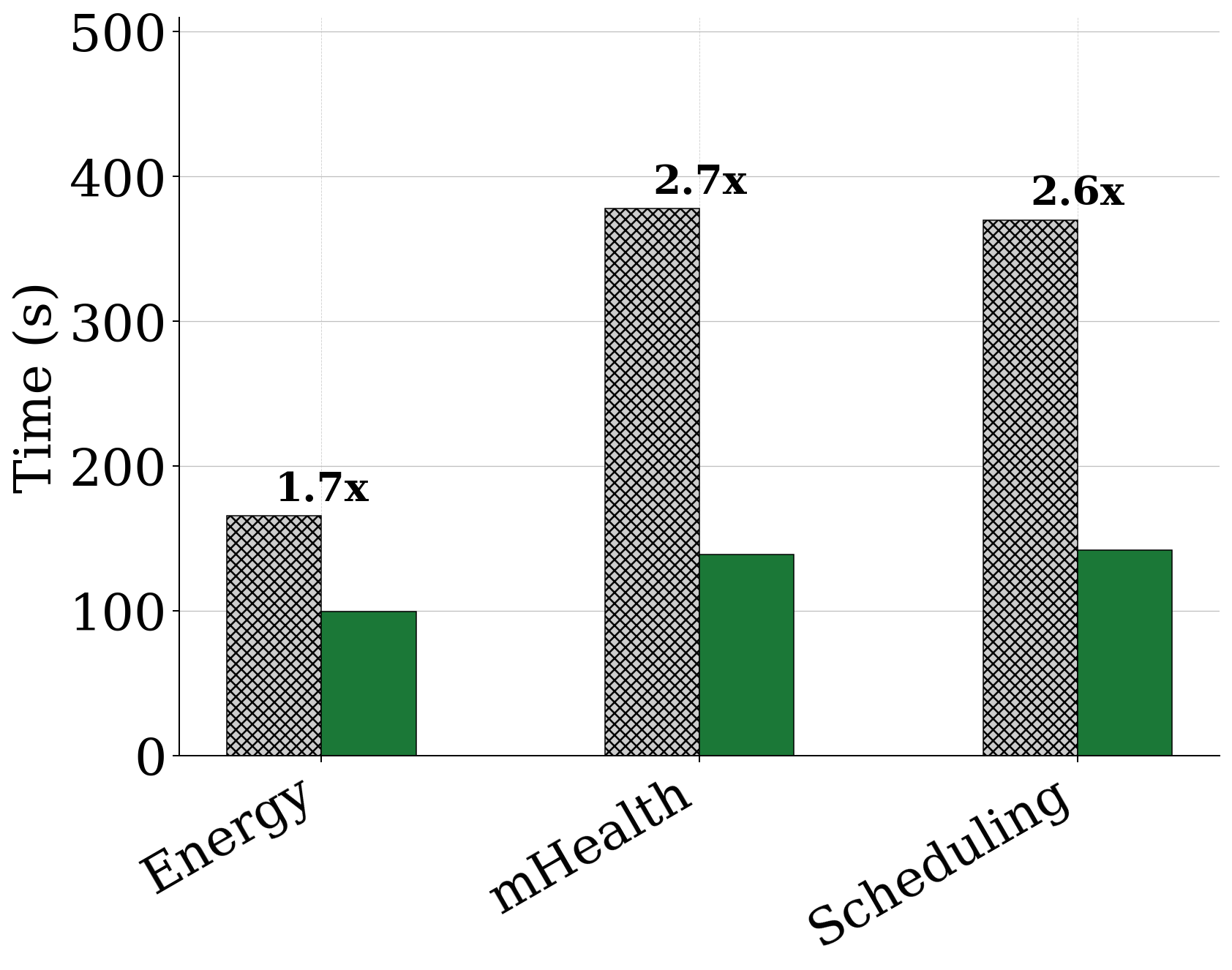}\vspace{-2mm}
        \subcaption{ABY3 (\texttt{bm-LAN})}
        \label{fig:tva-3pc-lan}
    \end{minipage}
    \hfill
    \begin{minipage}[b]{0.24\textwidth}
      \centering
      \includegraphics[width=\textwidth]{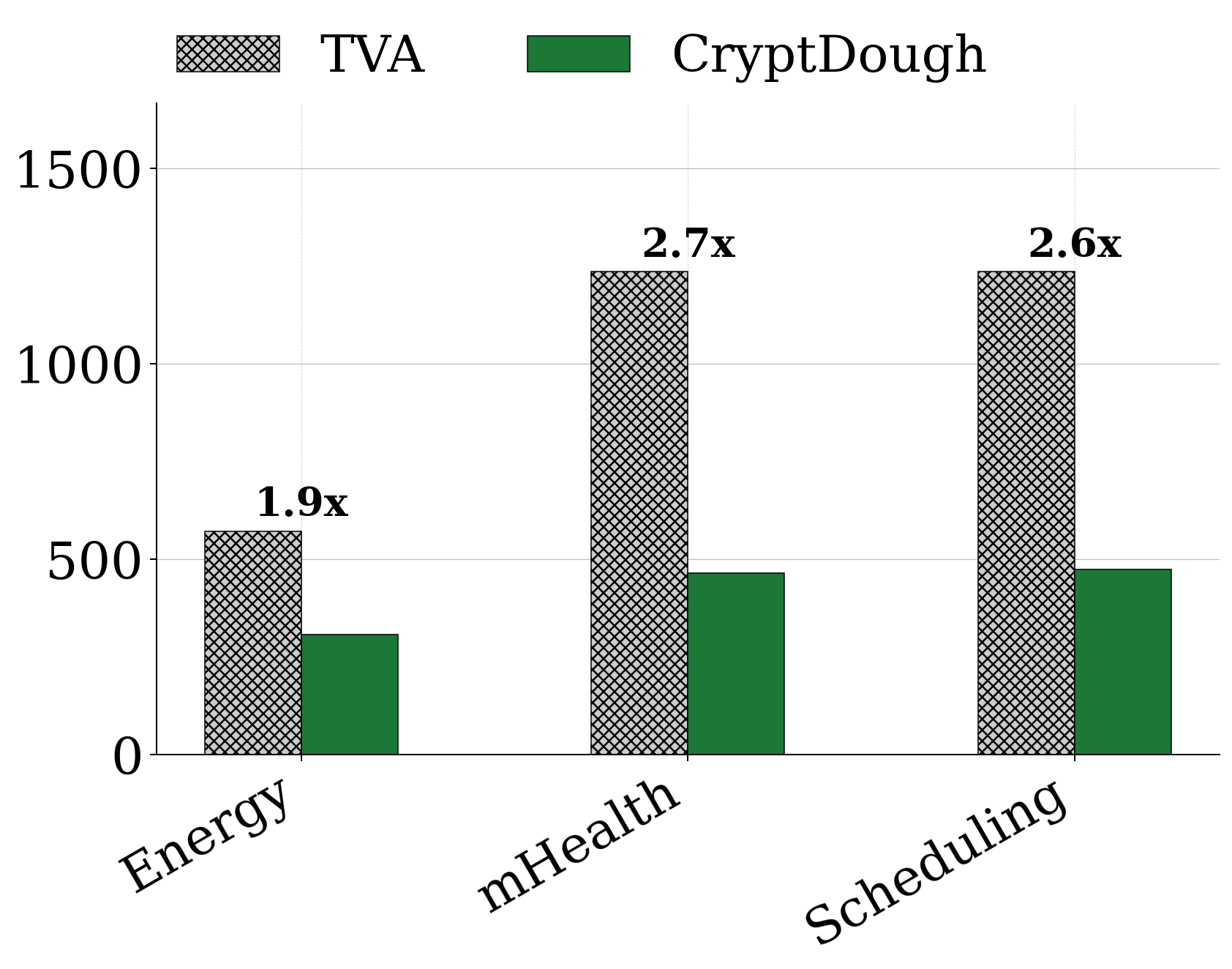}\vspace{-2mm}
      \subcaption{ABY3 (\texttt{bm-WAN})}
      \label{fig:tva-3pc-wan}
    \end{minipage}
    \hfill
    \begin{minipage}[b]{0.24\textwidth}
        \centering
        \includegraphics[width=\textwidth]{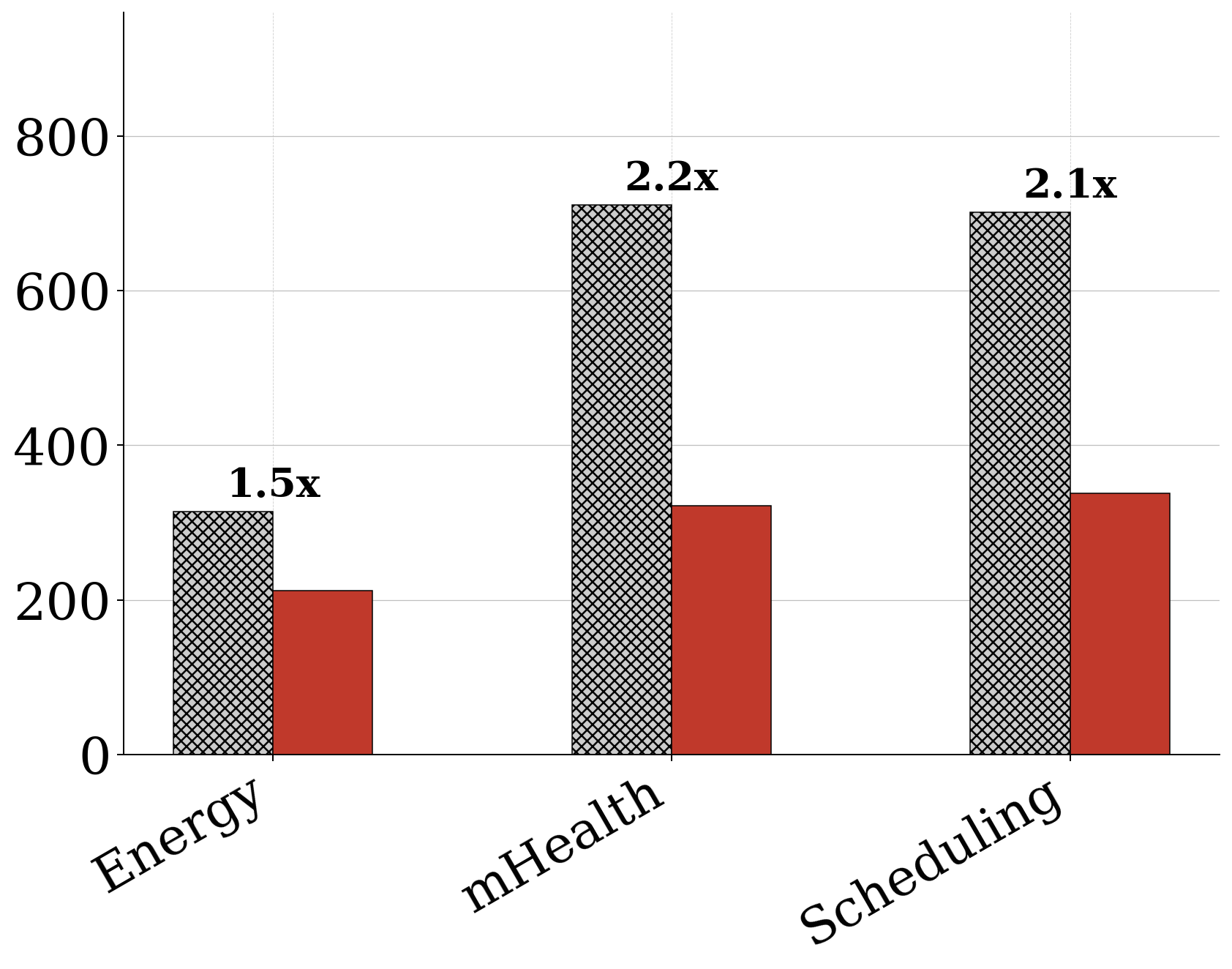}\vspace{-2mm}
        \subcaption{F4 (\texttt{bm-LAN})}
        \label{fig:tva-4pc-lan}
    \end{minipage}
    \hfill
    \begin{minipage}[b]{0.24\textwidth}
        \centering
        \includegraphics[width=\textwidth]{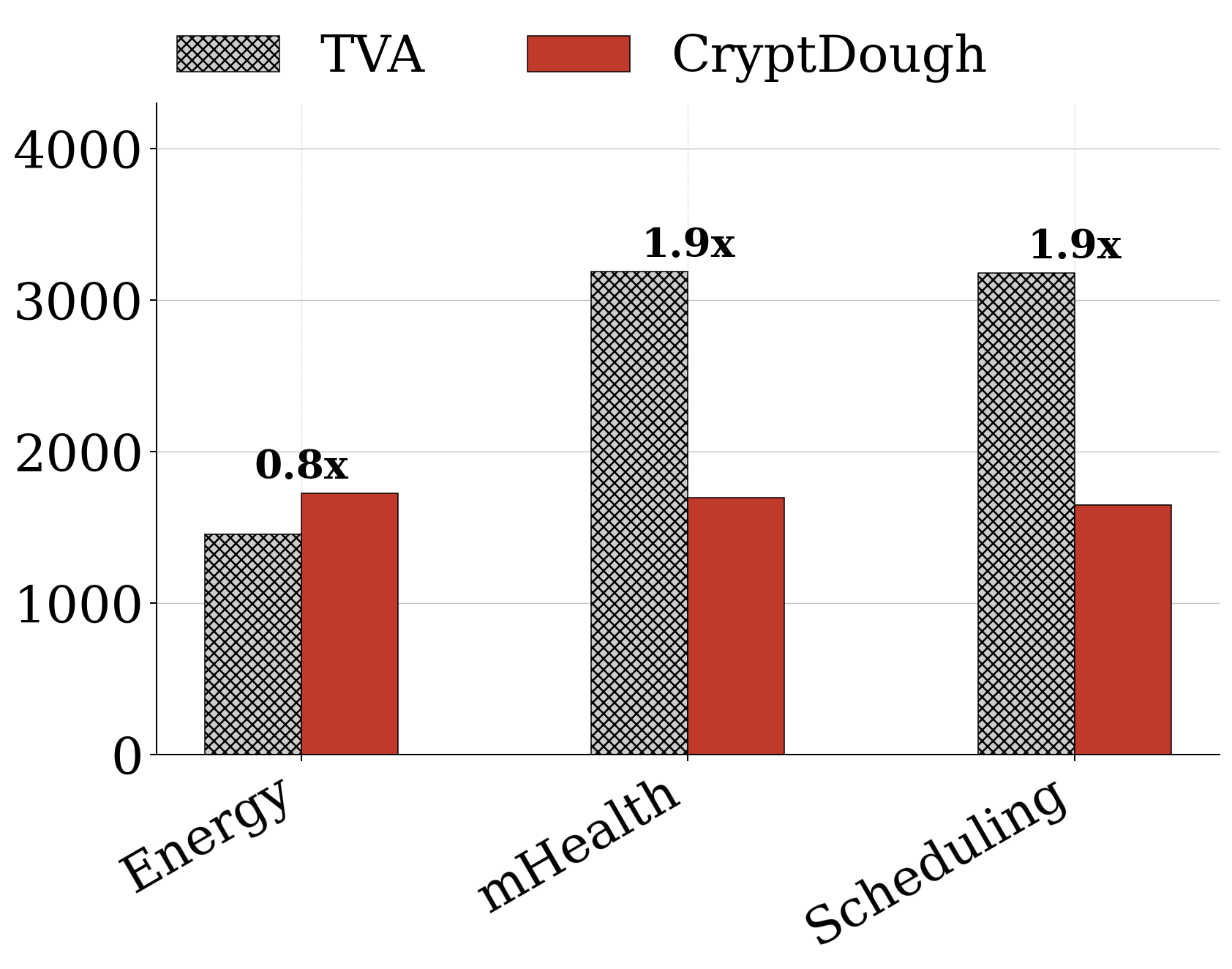}\vspace{-2mm}
        \subcaption{F4 (\texttt{bm-WAN})}
        \label{fig:tva-4pc-wan}
    \end{minipage}\vspace{-4mm}
    \caption{\camera{Comparison with TVA on all real-world applications used in the TVA paper}}\label{fig:comparison-tva}
\end{figure*}

In this section, we compare \ours with five state-of-the-art systems on the workloads they support, as summarized in Table~\ref{tab:competitors}. We always compare systems using protocols that fall under the same threat model; however, we note that some protocols are more efficient than those in \ours, which favors the baseline. For relational and time series analytics, we use ORQ~\cite{baum2025orq} and TVA~\cite{faisal2023tva}, respectively. For ML workloads, we compare with Pigeon~\cite{harth2025pigeon} and Piranha~\cite{watson2022piranha}, two recent systems for secure ML inference. The Pigeon paper reports that Pigeon-CPU outperforms Piranha for honest-majority MPC, which we have validated independently. Hence, we compare with Piranha only in the semi-honest dishonest-majority setting, which Pigeon does not support. For malicious security with dishonest majority, we compare with the MP-SPDZ MPC compiler~\cite{keller2020mp}. While MP-SPDZ does not provide built-in support for relational or time series analytics, it offers an API for CNN inference under SPDZ2k~\cite{mp-spdz-pytorch} and a custom AlexNet implementation that we use for  comparison.

\stitle{Relational analytics.}
We compare with ORQ on the 9 real-world queries in the ORQ paper: \emph{Aspirin},\emph{C. Diff. }\emph{Password}, \emph{Credit Score}, \emph{Comorbidity}, \emph{SecQ2}, \emph{Market Share}, \emph{SYan}, and \emph{Patients}. We set the input size to scale factor 1 ($\approx 5\mathrm{M}$ rows per table), as in the ORQ paper, and use oblivious quicksort for both systems.  We run experiments with two of the three protocols ORQ supports, since a recent paper~\cite{cryptoeprint:2026/234} identified an attack on the Fantastic Four (F4) protocol, and ORQ's security fix was in progress at the time of writing~\cite{orq-commit-4pc} (to the best of our understanding, \ours's implementation of F4 is not vulnerable to this attack). We use ORQ's default configuration for both systems: 16 threads with 4 connections in LAN and 16 connections in WAN. 
Fig.~\ref{fig:comparison-orq} shows that \ours achieves an average speedup of $1.5\times$~over~ORQ.

\stitle{Time series analytics.}
For the comparison with TVA, we use the three real-world applications in the TVA paper~\cite[\S6.2]{faisal2023tva}: \emph{monitoring energy consumption}, \emph{mobile health analytics}, and \emph{job scheduling optimization}. We use $2^{22}$ rows per time series, which is the largest input TVA reports results for.  For a fair comparison, we configure \ours to use bitonic sort and $32$-bit shares (with $64$-bit timestamps), like TVA, and set the number of threads to 16 for both systems (TVA's default). We show results for both TVA protocols, ABY3, and F4 (as far as we know, TVA's F4 implementation is not vulnerable to the attack~\cite{cryptoeprint:2026/234}). The results in Fig.~\ref{fig:comparison-tva} demonstrate that \ours outperforms TVA in all but one configuration, offering up to $2.7\times$ lower latency.

\stitle{ML inference: \ours vs. Pigeon.} For the comparison with Pigeon, we evaluate inference using (i) AlexNet on CIFAR-10, and (ii) VGG16 on CIFAR-10 and ImageNet. We use 192 images (the largest batch size reported in the Pigeon paper) and we configure both systems with 24 processes. Pigeon is a highly-optimized secure ML system that can leverage AVX instructions and a communication-efficient asymmetric 3-party protocol~\cite{harth2025high}. It further provides implementations of custom circuits for boolean addition to accelerate ReLU layers. We enable all of these optimizations in our experiments.

Table~\ref{tab:comparison-pigeon} shows the results. Despite being a general-purpose framework without any ML-specific optimizations or protocols, \ours is competitive with Pigeon and even outperforms it in LAN. In a high-RTT setting, Pigeon's custom ReLU gives it an advantage, as it significantly reduces the amount of communication required.

\stitle{ML inference: \ours vs. Piranha.} Piranha is a GPU-only inference framework. Since our bare-metal cluster does not have GPUs, we run Piranha in \texttt{AWS-LAN} and \texttt{AWS-WAN} on instances having the same GPU type as the one used for Piranha in the Pigeon paper~\cite{harth2025pigeon}.
We run the corresponding \ours experiments in \texttt{bm-LAN} and \texttt{bm-WAN}. Since Piranha cannot leverage CPU multi-threading, we configure \ours to only use one thread. We also use the same models and datasets as in the previous experiment, but we restrict the input to 8 images for VGG16 on ImagetNet because Piranha runs out of memory with a larger input size. Table~\ref{tab:comparison-piranha} shows that, even with a single thread, \ours outperforms Piranha, in all but one case, achieving up to $2.2\times$ faster inference.

\begin{figure*}[t]
    \centering
    \begin{minipage}[b]{0.24\textwidth}
        \centering
        \includegraphics[width=\textwidth]{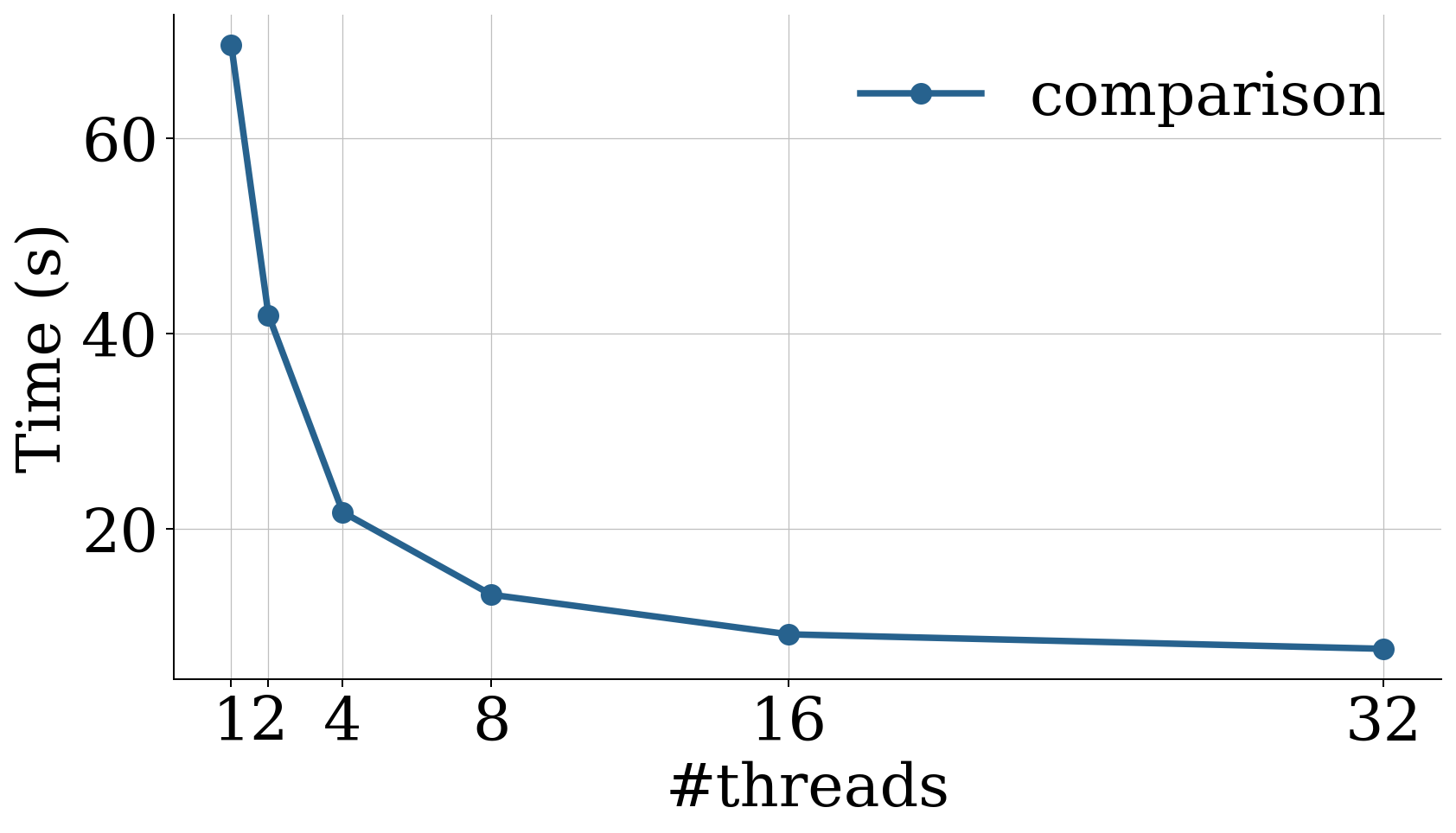}\vspace{-2mm}
        \subcaption{Comparison}
        \label{fig:gr-scalability}
    \end{minipage}
    \hfill
    \begin{minipage}[b]{0.24\textwidth}
        \centering
        \includegraphics[width=\textwidth]{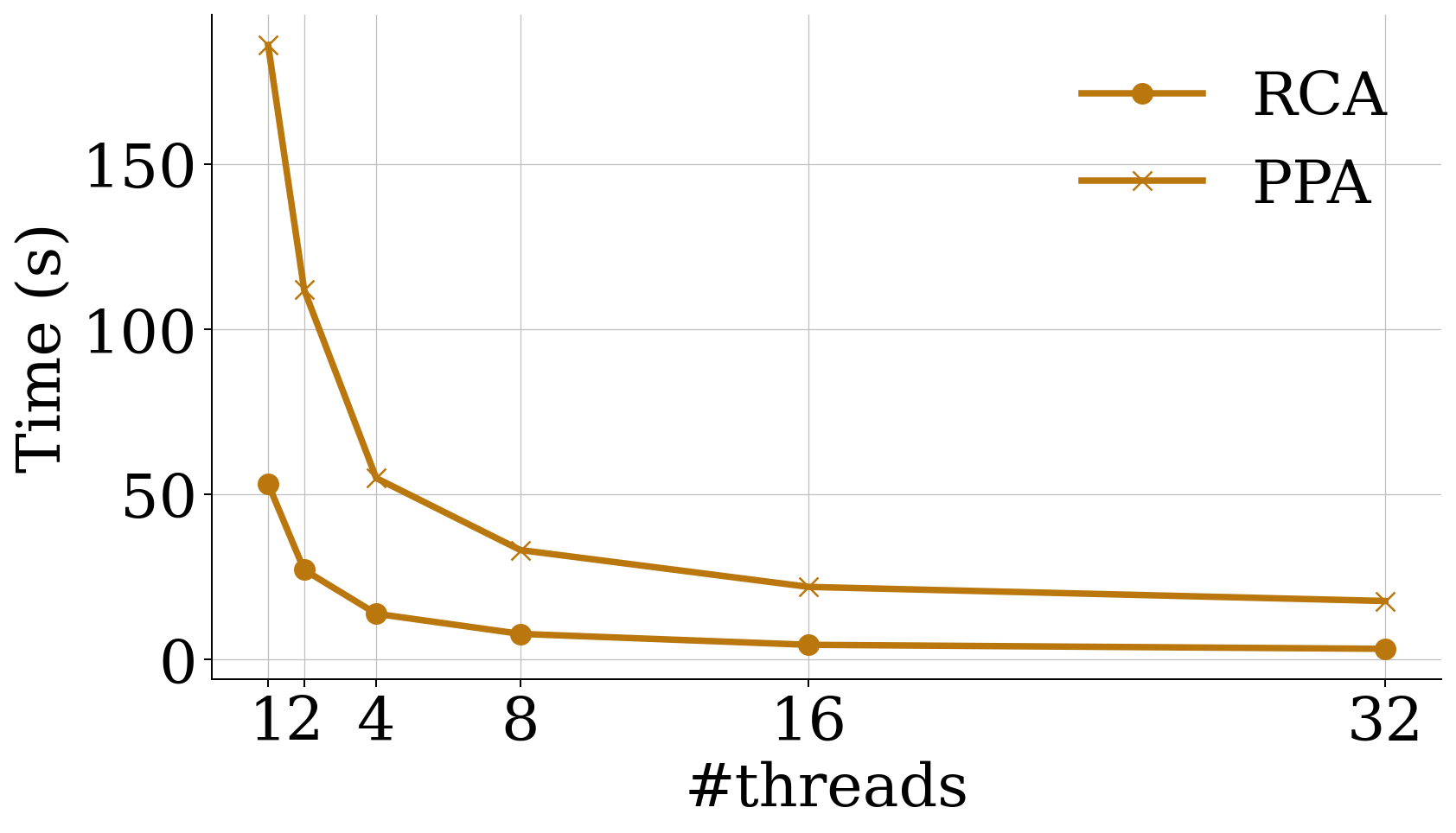}\vspace{-2mm}
        \subcaption{RCA/PPA}
        \label{fig:rca-ppa-scalability}
    \end{minipage}
    \hfill
    \begin{minipage}[b]{0.24\textwidth}
        \centering
        \includegraphics[width=\textwidth]{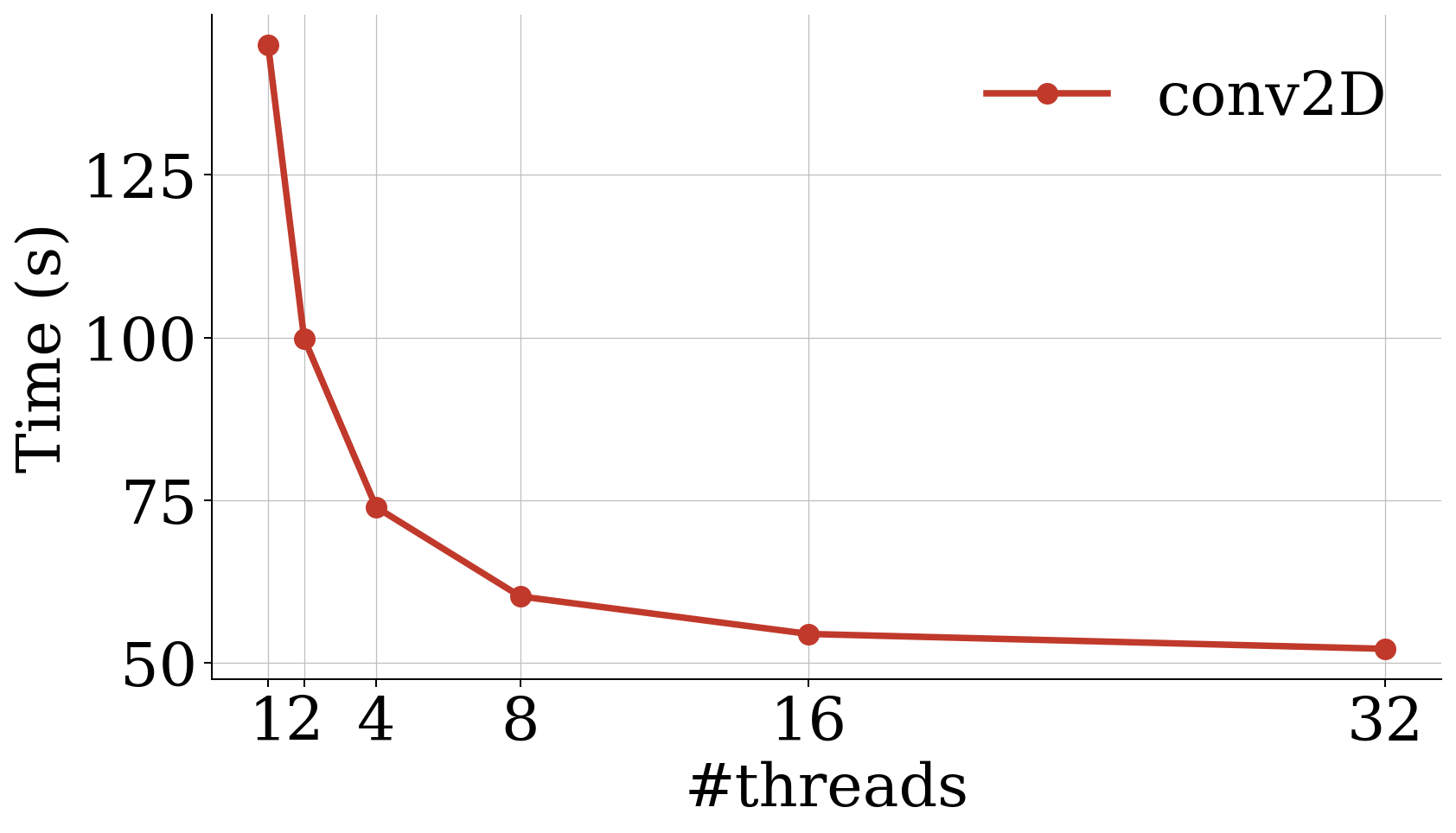}\vspace{-2mm}
        \subcaption{Conv2D}
        \label{fig:conv2d-scalability}
    \end{minipage}
    \hfill
    \begin{minipage}[b]{0.24\textwidth}
        \centering
        \includegraphics[width=\textwidth]{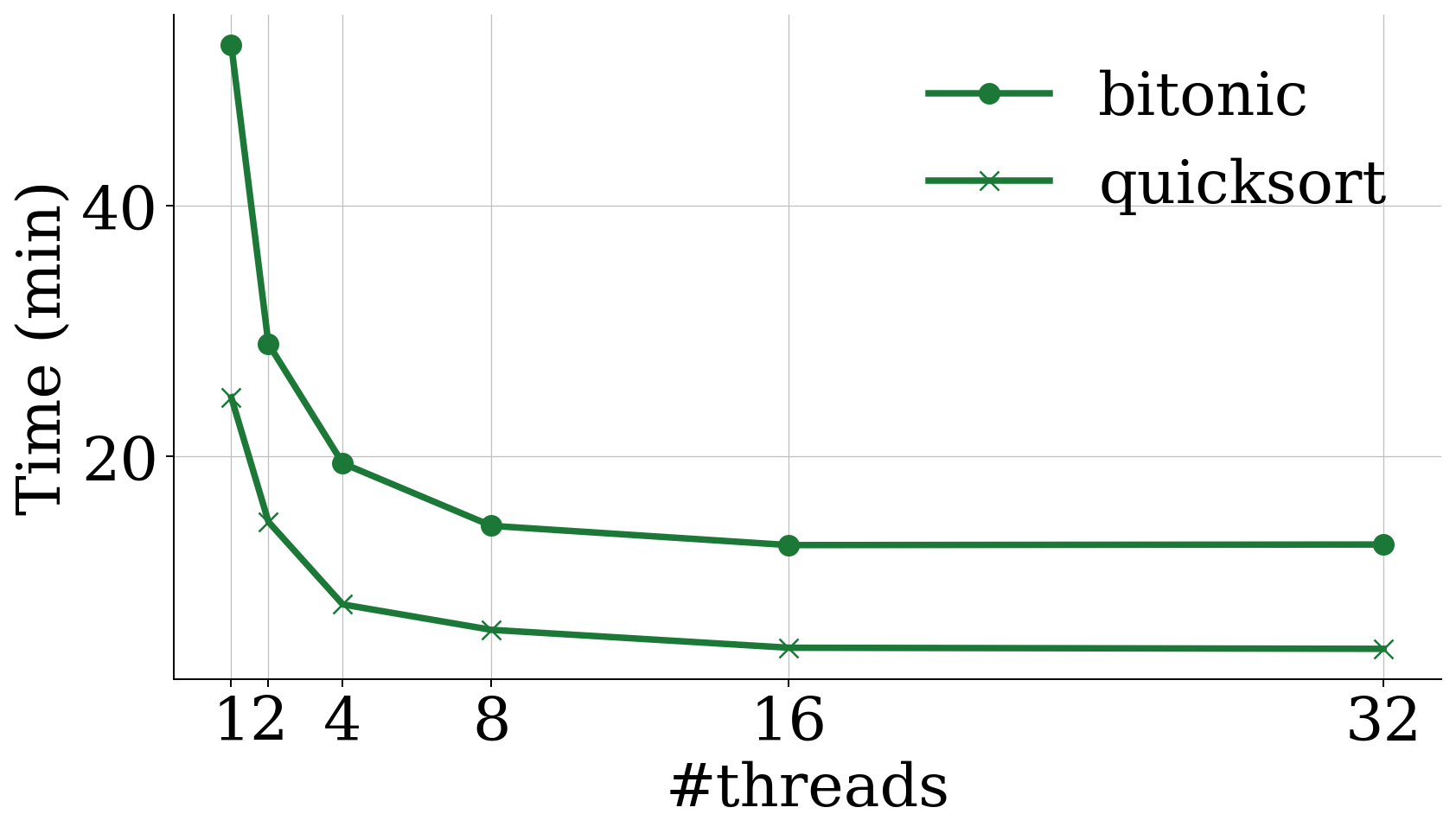}\vspace{-2mm}
        \subcaption{Sorting}
        \label{fig:sorting-scalability}
    \end{minipage}\vspace{-3mm}
    \caption{\camera{\ours's performance as we increase the number of worker threads per party~(ABY3, \texttt{bm-LAN})}}\label{fig:scalability}
\end{figure*}

\begingroup
\setlength{\tabcolsep}{3pt}
\begin{table}[t]
\centering\small
\begin{tabular}{clccc}
 & & {\bf AlexNet} & {\bf VGG16} & {\bf VGG16} \\
 & & {\bf (CIFAR-10)} & {\bf (CIFAR-10)} & {\bf (ImageNet)} \\
\hline\hline
\multirow{3}{*}{\rotatebox[origin=c]{90}{\bf LAN}}
 & {\bf Pigeon} & 0.49\,s      & 1.13\,s       & 52.3\,s \\
 & {\bf \ours} & {\bf 0.17\,s} & {\bf 1.11\,s} & {\bf 51.0\,s} \\
 & {Speedup} & $2.9\times$  & $1.0\times$   & $1.0\times$ \\
\hline
\multirow{3}{*}{\rotatebox[origin=c]{90}{\bf WAN}}
 & {\bf Pigeon} & 2.68\,s      & 5.52\,s       & 183.5\,s \\
 & {\bf \ours} & 3.92\,s      & 15.3\,s       & 219.6\,s \\
 & {Speedup} & $0.7\times$  & $0.4\times$   & $0.8\times$ \\
\end{tabular}
\caption{\camera{Comparison with Pigeon (Trio vs. ABY3)}}\label{tab:comparison-pigeon}\vspace{-3mm}
\end{table}
\endgroup

\begingroup
\setlength{\tabcolsep}{3pt}
\begin{table}[t]
\centering\small
\begin{tabular}{clccc}
 & & {\bf AlexNet} & {\bf VGG16} & {\bf VGG16} \\
 & & {\bf (CIFAR-10)} & {\bf (CIFAR-10)} & {\bf (ImageNet)} \\
\hline\hline
\multirow{3}{*}{\rotatebox[origin=c]{90}{\bf LAN}}
 & {\bf Piranha} & 0.57\,s       & 31.6\,s       & 47.3\,s \\
 & {\bf \ours}  & {\bf 0.46\,s} & {\bf 14.2\,s} & {\bf 30.1\,s} \\
 & {Speedup}  & $1.2\times$   & $2.2\times$   & $1.6\times$ \\
\hline
\multirow{3}{*}{\rotatebox[origin=c]{90}{\bf WAN}}
 & {\bf Piranha} & 3.58\,s       & 80.2\,s       & 164.4\,s \\
 & {\bf \ours}  & 9.20\,s       & {\bf 56.2\,s} & {\bf 92.3\,s} \\
 & {Speedup}  & $0.4\times$   & $1.4\times$   & $1.8\times$ \\
\end{tabular}
\caption{\camera{Comparison with Piranha (SecureML vs. ABY)}}\label{tab:comparison-piranha}\vspace{-3mm}
\end{table}
\endgroup

\stitle{ML inference: \ours vs. MP-SPDZ.} Table~\ref{tab:comparison-mpspdz} shows the comparison with MP-SPDZ. For this experiment, we run the custom AlexNet architecture provided in the public MP-SPDZ repository~\cite{mp-spdz-alexnet}, and we implement the same architecture in \ours. We configure both systems with 16 threads, no preprocessing, and an input size of 192 images. \ours outperforms  MP-SPDZ in both LAN and WAN; for WAN it achieves $4.7\times$ lower inference time, demonstrating state-of-the-art performance also under the strongest MPC threat model, which is  not supported by any of the above systems.

\begingroup
\setlength{\tabcolsep}{4pt}
\begin{table}[t]
\centering\small
\begin{tabular}{cccc}
 & \textbf{MP-SPDZ} & \textbf{\ours} & Speedup \\
\hline\hline
\textbf{LAN} & 28.45\,s & {\bf 24.07\,s} & $1.2\times$ \\
\hline
\textbf{WAN} & 581.9\,s & {\bf 124.7\,s} & $4.7\times$ \\
\end{tabular}
\caption{\camera{Comparison with MP-SPDZ (SPDZ2k) on AlexNet~\cite{mp-spdz-alexnet}}}\label{tab:comparison-mpspdz}\vspace{-3mm}
\end{table}
\endgroup

\subsection{Evaluating \ours's parallelization}\label{sec:exp-scalability}
Next, we assess how well the \ours runtime can automatically parallelize low-level MPC functionalities and high-level secure operators. To this end, we evaluate four secure primitives that serve as building blocks for \ours's analytics libraries: comparison, a Ripple-Carry Adder (RCA), a Parallel-Prefix Adder (PPA), and convolution. We use vectors with $2^{28}$ elements for comparison and adders, and 256 instances of the largest VGG16-ImageNet Conv2D layer for convolution. Fig.~\ref{fig:gr-scalability}, \ref{fig:rca-ppa-scalability}, and \ref{fig:conv2d-scalability} show the execution time as we increase the number of threads from 1 to 32, for ABY3  in \texttt{bm-LAN}. Experiments with other \ours protocols are given in Appendix~\ref{apdx:micro}.

Comparison and addition scale almost perfectly with up to 8 threads, achieving $\approx 2\times$ speedup at every step. Increasing the number of threads further amplifies the overheads of runtime coordination and communication, though, \ours still achieves $\approx 1.5\times$ speedup going from 16 to 32 threads. Convolution is more challenging to scale, as the ratio of communication to computation is high, and overlapping provides only modest benefits: \ours achieves between $1.45\times$ and $1.1\times$ speedup when doubling the number of threads.

To understand the impact of primitive composition on scaling, we run bitonic sort and quicksort on a vector of $2^{26}$ elements. These are two high-level oblivious operators that make black-box use of lower-level functionalities and form the basis for many of \ours's relational and time series operations. The results in Fig.~\ref{fig:sorting-scalability} show that the \ours runtime can effectively parallelize both, even though the library developer expressing the oblivious control flow of these sorting networks writes single-threaded code.

\section{Related work}\label{sec:related}

\stitle{MPC systems for analytics.}
In the ML space, existing MPC-based systems support a range of tasks, including linear model fitting~\cite{cerebro, mohassel2017secureml, scalable-linear-reg-mpc, helen, blaze-ml}, neural network inference~\cite{DBLP:conf/uss/MishraLSZP20,gazelle,DBLP:conf/sp/PangZMZS24}, and even training~\cite{watson2022piranha, harth2025pigeon, crypten, TanKTW21, sefspu, orca, wagh2021falcon,adam-in-private}.
For relational analytics, systems like ORQ~\cite{baum2025orq}, Conclave~\cite{Volgushev2019Conclave}, and others~\cite{Poddar2021Senate, liagouris2023secrecy, Bater2017SMCQL, FengScape2022, Wang2021Secure, fang2024secretflow, bater2018shrinkwrap, bater2020saqe, DBLP:conf/ndss/ChowLS09, 10.1145/3658644.3690314} support a broad class of database queries on tabular data.
The time series and graph analytics space are comparatively less explored; Waldo~\cite{waldo} and TVA~\cite{faisal2023tva} support time series and some basic relational operations (filters and aggregations), 
while GraphSC~\cite{graphsc} and others~\cite{graphitti, 10.1145/3460120.3484560} target graph analysis that \ours does not currently support. 

Unlike \ours, all these systems are designed for a single workload type---ML, relational, time series, or graph analytics---and cannot naturally support mixed pipelines that arise in practice. Moreover, the vast majority of these systems is tailored to a single threat model, further limiting their applicability.  \ours is the first MPC-based analytics system to report results across the full threat model spectrum, encompassing both semi-honest and malicious adversaries in honest- and dishonest-majority settings. 

\stitle{MPC compilers.}
Frameworks like MP-SPDZ~\cite{keller2020mp}, \camera{MOTION \cite{motion}}, and others~\cite{10.1145/3453483.3454074, DBLP:conf/sp/HastingsHNZ19, picco, silph, wysteria, scale-mamba, ezmpc, emp-toolkit, oblivm, fairplay} have made MPC
more accessible by allowing users to write arbitrary programs in high-level procedural languages (e.g., subsets of Python or C) and compile them into secure MPC circuits. This line of work focuses on cost-based compiler optimizations, such as automatic vectorization, expression rewriting~\cite{kerschbaum-rewriting}, or \camera{mixing different MPC protocols in the same user program~\cite{motion, costco}}.
While \ours could incorporate such techniques to optimize UDFs, our approach is fundamentally different.
\ours does not allow users to write arbitrary programs; in contrast, it provides vectorized, data-parallel operators for analytics that can be composed arbitrarily to create mixed workflows like the one in Fig.~\ref{fig:mixed-pipeline}. Expressing such pipelines in the above frameworks is technically possible, but would have the same hurdles we discussed in \S\ref{sec:workloads}. \camera{This holds even for systems like MP-SDPZ and MOTION that provide built-in support for CNN inference, since users would still need to implement the relational and time series operators directly from low-level primitives.}

\stitle{Custom MPC solutions.}
There are also several MPC solutions that target specific use cases and workloads, including aggregate statisitcs~\cite{DBLP:conf/ccs/BonawitzIKMMPRS17, sepia, DBLP:conf/eurosp/IonKNPSS0SY20}, private matching~\cite{cryptoeprint:2020:599, DBLP:journals/popets/MourisMTSBC24}, and heavy hitters~\cite{whisper, prio}, among others. 
Splinter~\cite{splinter} uses function secret sharing to evaluate private queries on public data, while
Flock~\cite{kaviani2024flock} targets secure key management. 

\balance

\section{Conclusion}

We presented \ours, the first unified analytics engine for secure multiparty computation. 
\ours provides built-in support for relational queries, time series computations, ML inference tasks, and their composition, generalizing the functionality of prior solutions, without compromising security or performance.
We believe \ours will be a useful tool towards pushing MPC beyond current niche applications.


\begin{acks}

\camera{This paper will appear at SOSP 2026.  We thank our shepherd and the anonymous reviewers for their thoughtful comments that substantially improved the paper, and the artifact evaluators for reproducing the results of this work. We also thank our collaborators Nicole Spartano, Lisa Quintiliani, Edwin Boudreaux, Ola Ozernov-Palchik, Rhoda Au, the Boston Women's Workforce Council (BWWC), the Massachusetts Technology Collaborative (MassTech),  the Massachusetts Executive Office of Labor and Workforce Development (EOLWD), Bosch GmbH, and Red Hat Inc for providing real use cases that have influenced \ours's design.  We are also grateful to the following individuals who contributed by providing code reviews and useful feedback on system design: Jerry Zhang,  Nitin Mathai,  Eli Baum, Thomas Unger, Islam Faisal, and Mohamed Oscar.
This paper is based upon work supported by NSF awards No.\ 2209194, 2541869, and 2613424; by REU supplement awards No.\ 2432612 and 2326580; and by an Amazon Research Award. Initial development and prototyping was done with computing resources provided by the Chameleon~\cite{chameleon} and CloudLab~\cite{cloudlab} research testbeds supported by the National Science Foundation.}

\end{acks}

\bibliographystyle{ACM-Reference-Format}
\bibliography{references}

\clearpage
\appendix

\nobalance

\section{Security Analysis}\label{apdx:security-analysis}

\subsection{Security of composition}
\camera{ \ours uses the functionalities of its underlying MPC protocols in a black-box manner, supporting mixed-mode operations and conversions. \ours is designed so that each layer directly inherits the security guarantees of the layer below it, and ultimately of the underlying MPC protocol. Concretely, our higher-level operators are built from the secure primitives exposed by the protocol layer (e.g., addition, multiplication, XOR-gate, and AND-gate) and listed in Table~\ref{tab:functionalities}. Higher-level operators invoke these primitives without inspecting or branching on any secret value. Their control flow is data-independent, so an operator's execution and access pattern reveal nothing about its inputs, outputs, or intermediate data. Primitives are instantiated following the constructions of Demmler et al.~\cite{aby}, Araki et al.~\cite{araki2016high,aby3}, Dalskov et al.~\cite{dalskov2021fantastic,cryptoeprint:2026/234}, and Cramer et al.~\cite{spdz-2k}.}

\camera{Thanks to this black-box model and data-independent composition, the security of \ours's operators and analytics layers follows in the standard arithmetic black-box (ABB) model: any protocol that securely realizes the primitive functionalities can be composed, and the resulting higher-level protocol is secure by the standard composition argument. It thereby inherits the privacy and correctness guarantees---as well as the threat model (semi-honest or malicious)---of the instantiated primitives. Prior work~\cite{faisal2023tva,DBLP:conf/crypto/EscuderoGKRS20, DBLP:conf/acns/KellerORSSV17,DBLP:conf/ccs/0001RRSS16} has used a similar argument to prove system security.}

\subsection{Security assumptions}

\camera{Like all MPC systems for analytics,  \ours has the entire codebase in the trusted computing base (TCB); that is, a flaw in any part of the implementation can foil the MPC security guarantees. However, \ours does not rely on trusted compute or storage; in other words, the security guarantees hold even if an administrator observes the state of a server during and between computations. As a result, \ours can be deployed in a typical outsourced setting where data owners secret-share their data to untrusted computing parties. The network layer is also considered untrusted, so \ours needs to establish authenticated, encrypted channels between servers using TLS.}

\camera{\ours's fully oblivious implementation across all layers of the stack eliminates known timing-based side channels. We make no explicit attempt to prevent other forms of side channels (e.g., cache or power). Because MPC programs are oblivious, side-channel leakage at a small number of parties is naturally mitigated: the computation behaves analogously to a masking countermeasure, revealing nothing about secret values from any single party's execution. Formal verification of the \ours software is an exciting direction for future work; the current modular architecture would greatly facilitate this process.}

\section{Example CNN model implementation}\label{apdx:vgg16}

Listing~\ref{lst:vgg16-implemntation} shows the implementation of the fine-tuned VGG16 model~\cite{xray-vgg16} that we use for X-ray classification in the multi-workload pipeline of Fig.~\ref{fig:mixed-pipeline} and the experiments of \S\ref{sec:exp-mixed}. 

\begin{lstlisting}[language=C++, caption={VGG16 model implementation with \ours's API}, numbers=left, label={lst:vgg16-implemntation}, escapechar=|]
Model<T, SecureMatrix, Engine> VGG16(Engine& engine) {
  const size_t precision = 16; // Fixed-point precision
  // Model initialization
  Model<T, SecureMatrix, Engine> m(engine, precision);
    
     	 /*** Construct VGG16 model layers ***/  

  // HW(X,Y) defines layer dimensions (height,width)
  m.conv2D(HW(224,224),3,64,HW(3,3),HW(1,1),HW(1,1));
  m.reLU(3211264);
  m.conv2D(HW(224,224),64,64,HW(3,3),HW(1,1),HW(1,1));
  m.avgPool(HW(224,224),64,HW(2,2),HW(2,2),HW(0,0));
  m.reLU(802816);
  m.conv2D(HW(112,112),64,128,HW(3,3),HW(1,1),HW(1,1));
  m.reLU(1605632);
  m.conv2D(HW(112,112),128,128,HW(3,3),HW(1,1),HW(1,1));
  m.avgPool(HW(112,112),128,HW(2,2),HW(2,2),HW(0,0));
  m.reLU(401408);
  m.conv2D(HW(56,56),128,256,HW(3,3),HW(1,1),HW(1,1));
  m.reLU(802816);
  m.conv2D(HW(56,56),256,256,HW(3,3),HW(1,1),HW(1,1));
  m.reLU(802816);
  m.conv2D(HW(56,56),256,256,HW(3,3),HW(1,1),HW(1,1));
  m.avgPool(HW(56,56),256,HW(2,2),HW(2,2),HW(0,0));
  m.reLU(200704);
  m.conv2D(HW(28,28),256,512,HW(3,3),HW(1,1),HW(1,1));
  m.reLU(401408);
  m.conv2D(HW(28,28),512,512,HW(3,3),HW(1,1),HW(1,1));
  m.reLU(401408);
  m.conv2D(HW(28,28),512,512,HW(3,3),HW(1,1),HW(1,1));
  m.avgPool(HW(28,28),512,HW(2,2),HW(2,2),HW(0,0));
  m.reLU(100352);
  m.conv2D(HW(14,14),512,512,HW(3,3),HW(1,1),HW(1,1));
  m.reLU(100352);
  m.conv2D(HW(14,14),512,512,HW(3,3),HW(1,1),HW(1,1));
  m.reLU(100352);
  m.conv2D(HW(14,14),512,512,HW(3,3),HW(1,1),HW(1,1));
  m.avgPool(HW(14,14),512,HW(2,2),HW(2,2),HW(0,0));
  m.reLU(25088);

  // Task-specific layers for Xray classification (|\S\ref{sec:exp-mixed}|)
  m.avgPool(HW(7,7),512,HW(2,2),HW(2,2),HW(0,0));
  m.reLU(512);
  m.fullyConnected(512, 256);
  m.reLU(256);
  m.fullyConnected(256, 256);
  m.reLU(256);
  m.fullyConnected(256, 3);
  m.reLU(3);
  
  return m; // Can be populated with pretrained weights
}
\end{lstlisting}

\section{\ours protocol layer interface}\label{apdx:interface}
Table~\ref{tab:functionalities} shows the interface that an applied cryptographer must implement to add a new MPC protocol to \ours.

\section{Custom parallelization example}\label{apdx:custom-example}

Listing~\ref{lst:custom-op} shows a baseline implementation of the oblivious control from Fig.~\ref{fig:vv-threads} using \ours's vectorized \texttt{cmp-swp()} primitive (\S\ref{sec:runtime}). Without virtual vectors, library developers would need to write custom code to handle operator-specific task-to-worker assignment (lines~\ref{lst:mt-start}-\ref{lst:mt-end}). Reusing low-level vectorized primitives, such as \texttt{cmp-swp()}, would also require ``packing'' data in temporary vectors (lines~\ref{lst:cp-start}-\ref{lst:cp-end}) and copying results back into the original vectors (lines~\ref{lst:cp-bk-start}-\ref{lst:cp-bk-end}). \ours's virtual vectors allow developers to express the same parallelization logic, without redundant copies, in three lines of code, as shown in Listing~\ref{lst:vv-example} of the main paper.\vspace{1mm}

\begin{lstlisting}[language=C++, caption={Example implementation of the oblivious control flow from Fig.~\ref{fig:vv-example} that showcases the challenges of modularity and parallelization in the absence of \ours's virtual vector abstractions.}, numbers=left, label={lst:custom-op}, escapechar=|]
// Worker's task
void w-task(BSharedVector<T>& v, int start, int end, 
						int d, int stride, int num_ops) {
	int n = v.size();
	BSharedVector tmp1(num_ops);
	BSharedVector tmp2(num_ops);
	// Copy comparison elements into temporary vectors
	for (int i=start,j=0; i<end && i+d<n; i+=stride,j++) {|\label{lst:cp-start}|
		tmp1[j] = v[i];
		tmp2[j] = v[i+d];
	}|\label{lst:cp-end}|
	// Apply secure compare-and-swap primitive (Listing|~\ref{lst:cmp-swp}|)
	|\textbf{cmp-swp}|(tmp1, tmp2);
	// Copy swapped elements back to the input vector
	for (int i=start,j=0; i<end && i+d<n; i+=stride,j++) {|\label{lst:cp-bk-start}|
		v[i] = tmp1[j];
		v[i+d] = tmp2[j];
	}|\label{lst:cp-bk-end}|
}

// Main thread
BSharedVector<int> v = ... // Input vector|\label{lst:mt-start}|
int n = v.size(), d=3, stride=2, t=NUM_THREADS;
int pairs = std:ceil((n-d)/stride);
int pairs_per_tid = std::ceil(pairs/t); 
std::vector<std::thread> threads;
for (int tid=0; tid<num_threads; tid++) {
	// Compute worker's partition
	int start = tid * pairs_per_tid * stride;
	int end = std::min(start + pairs_per_tid*stride, n-d);
	if (start < n-d) { // Assign task
		threads.emplace_back(w-task, std::ref(v),  
							 					 start, end, d, stride,
							 					 pairs_per_tid);
	}
	for (auto& thread : threads) {
		thread.join();
	}
}|\label{lst:mt-end}|
\end{lstlisting}

\begin{lstlisting}[language=C++, caption={C++ equivalent code for the simple pattern}, numbers=left, label={lst:simple-equivalence}, escapechar=|]
auto y = x.simple(offset, step, end);
int i = 0;
for(int f_i=offset; f_i<end; f_i+=step){
	/* f_i = offset + step * i */
	/* y[i] == x[f_i] */
	i++;
}
\end{lstlisting}

\begin{lstlisting}[language=C++, caption={C++ equivalent code for the repeated pattern}, numbers=left, label={lst:repeated-equivalence}, escapechar=|]
auto y = x.repeated(n);
int i = 0;
for(int f_i=0; f_i<x.size(); f_i++){
	for(int i1 = 0; i1<n; i1++){
		/* f_i = i/n */
		/* y[i] == x[f_i] */
		i++;
	}
}
\end{lstlisting}

\pagebreak

\section{Core access patterns in \ours}\label{apdx:access-patterns}
\begin{figure}[t]
  \centering
      \includegraphics[width=\linewidth]{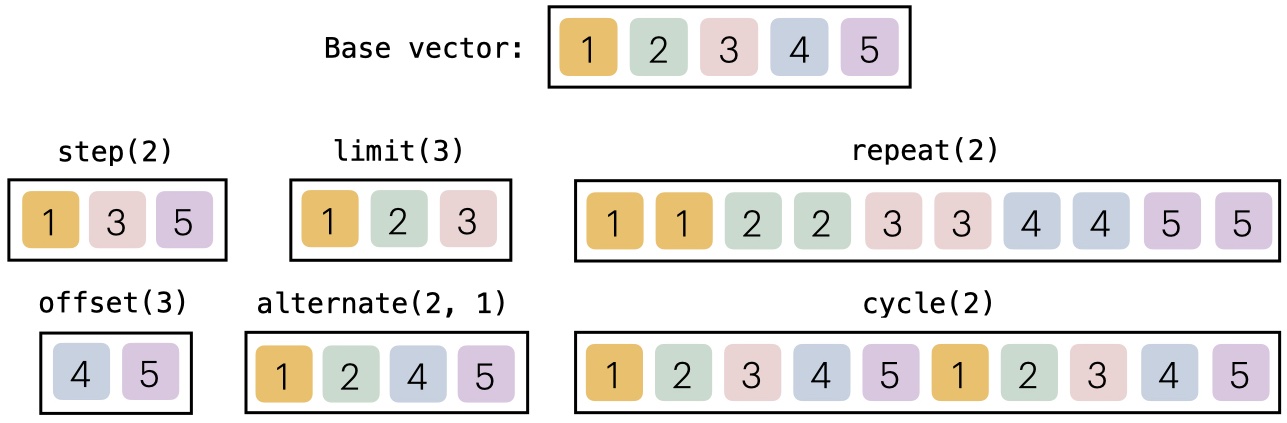}
  \caption{Examples of core data access patterns in \ours}\label{fig:patterns-example}
\end{figure}

Table~\ref{tab:access-patterns} summarizes the core data access patterns used in \ours, along with their mapping functions and equivalent C++ implementations.  Listing~\ref{lst:access-pattern-alias} shows aliases for common access patterns, and Fig.~\ref{fig:patterns-example} provides concrete examples. The core access patterns can be composed arbitrarily to express complex oblivious control flows that the system runtime can parallelize automatically, as explained in \S\ref{sec:virtual-vectors}.

\begin{table*}[ht]
\centering\small
\begin{tabular}{l|l}
\textbf{Functionality} & \textbf{Description}\\ 
\hline
\texttt{void \textbf{add\_a}(const EVector\& x, const EVector\& y, EVector\& z)} & Secure addition ($z = x + y$)\\ 
\hline
\texttt{void  \textbf{multiply\_a}(const EVector\& first, const EVector\& second, EVector\& result)} & Secure multiplication ($z = x \cdot y$) \\ 
\hline
\texttt{void \textbf{xor\_b}(const EVector\& x, const EVector\& y, EVector\& z)} & Secure bitwise XOR ($z = x \oplus y$) \\ 
\hline
\texttt{void \textbf{and\_b}(const EVector\& first, const EVector\& second, EVector\& result)} & Secure bitwise AND ($z = x \land y$) \\ 
\hline
\texttt{void \textbf{b2a}(const EVector\& in, EVector\& out)} & Boolean-to-Arithmetic conversion \\ 
\hline
\texttt{void \textbf{a2b}(const EVector\& in, EVector\& out)} & Arithmetic-to-Boolean conversion \\ 
\hline
\texttt{std::vector<EVector> \textbf{get\_shares\_a}(const Vector\& data)} & Create arithmetic shares \\ 
\hline
\texttt{std::vector<EVector> \textbf{get\_shares\_b}(const Vector\& data)} & Create boolean shares \\ 
\hline
\texttt{Vector \textbf{open\_shares\_a}(const EVector\& shares)} & Open arithmetic shares\\ 
\hline
\texttt{Vector \textbf{open\_shares\_b}(const EVector\& shares)} & Open boolean shares\\ 
\hline
\end{tabular}
\caption{Adding a new MPC protocol to \ours requires implementing a short list of low-level functionalities for creating, exchanging, and manipulating secret shares. All functionalities are vectorized by design, i.e., they are implemented on top of the \texttt{EVector} and \texttt{Vector} abstractions. User programs are compiled against  protocol functionalities and inherit their security guarantees.}\label{tab:functionalities}
\end{table*}

\begin{table*}[t]
\centering
\small
\begin{tabular}{l|p{6cm}|p{8cm}}
\textbf{Pattern} & \textbf{Mapping Function} & \textbf{Description} \\
\hline
Simple & $f(i) = offset + i \cdot step$\newline\newline \textbf{where} $0 \leq i < \lceil \frac{limit - offset}{step} \rceil$ & 
\itemize{
	\item {Iterates over elements with a \texttt{step}, starting from an optional \texttt{offset} and up to an optional \texttt{limit}.}
	\item{Useful for selecting subsets (e.g., adjacent elements).}
	\item{Equivalent to a \textbf{single for-loop} shown in Listing~\ref{lst:simple-equivalence}.}
	} \\

\hline
Repeated & $f(i) = \lfloor i / n \rfloor$\newline\newline \textbf{where} $0 \leq i < n \cdot |V|$ &
\itemize{
	\item{Repeats each element $n$ times consecutively, expanding the vector by a factor of $n$.}
	\item{Useful for operations between vectors and single values.}
	\item{Equivalent to an \textbf{outer for-loop} in nested for-loops,  as shown in Listing~\ref{lst:repeated-equivalence}.}
	}\\

\hline
Cyclic & $f(i) = i \mod |V|$\newline\newline \textbf{where} $0 \leq i < n \cdot |V|$ & 
\itemize{
	\item{Repeats the  vector $n$ times cyclically.}
	\item{Used in certain join operations.}
	\item{Equivalent to an \textbf{inner for-loop} in nested for-loops, as shown in Listing~\ref{lst:cyclic-equivalence}.}
}\\
\hline
Alternating & $f(i) = \lfloor i/m \rfloor \cdot (m+k) + (i \mod m)$\newline\newline \textbf{where} $0 \leq i < \lfloor |V|/(m+k) \rfloor \cdot m + min(m, |V| mod (m+k))$ & 
\itemize{
	\item{Alternates between including $m$ elements and excluding $k$ elements.}
	\item{used in sorting networks (e.g., bitonic sort).}
	\item{Equivalent to \textbf{two fused for-loops} in nested for-loops. For instance, a for-loop index is initialized using a higher level for-loop index, as shown in Listing~\ref{lst:alternating-equivalence}.}
}\\
\end{tabular}
\caption{A virtual vector $W$ is a restricted view of a base vector $V$ such that $W[i] = V[f(i)]$. This table summarizes the core access patterns provided by \ours to create virtual vectors, along with their mapping functions. The equivalent C++ implementations are given in Listings~\ref{lst:simple-equivalence}-\ref{lst:alternating-equivalence}. Virtual vectors can be defined on top of other virtual vectors to express more complex control flows. For example, the cyclic and simple patterns can be combined to cyclically traverse every other element of a base vector.}
\label{tab:access-patterns}
\end{table*}

\begin{lstlisting}[language=C++, caption={C++ equivalent code for the cyclic pattern}, numbers=left, label={lst:cyclic-equivalence}, escapechar=|]
auto y = x.cyclic(n);
int i = 0;
for(int i1=0; i1<n; i1++){
	for(int f_i = 0; f_i<x.size(); f_i++){
		/* f_i = i%n */
		/* y[i] == x[f_i] */
		i++;
	}
}
\end{lstlisting}

\begin{lstlisting}[language=C++, caption={C++ equivalent code for the alternating pattern}, numbers=left, label={lst:alternating-equivalence}, escapechar=|]
auto y = x.alternating(m, k);
int i = 0;
for(int i1=0; i1<x.size(); i1+=(k+m)){
	for(int f_i=i1; f_i<i1+k && f_i<x.size(); ++f_i){
		/* f_i = (i/m) * (m+k)+ i%m */
		/* y[i] == x[f_i] */
		i++;
	}
}
\end{lstlisting}

\begin{lstlisting}[language=C++, caption={Access patterns aliases used in \ours}, numbers=left, label={lst:access-pattern-alias}, escapechar=|]
x.offset(i) |$\Leftrightarrow$| x.shift(i,1,x.size()); 
x.step(k) |$\Leftrightarrow$| x.shift(0,k,x.size());
x.limit(s) |$\Leftrightarrow$| x.shift(0,1,s);
x.slice(i,s) |$\Leftrightarrow$| x.shift(i,1,s);
x.limit(s).offset(i).step(k) |$\Leftrightarrow$| x.shift(i,k,s);
\end{lstlisting}

\section{Additional experiments}\label{apdx:micro}

\subsection{Scalability of core primitives}
\begin{figure*}[t]
	\centering
	\begin{minipage}[b]{0.24\textwidth}
		\centering
		\includegraphics[width=\textwidth]{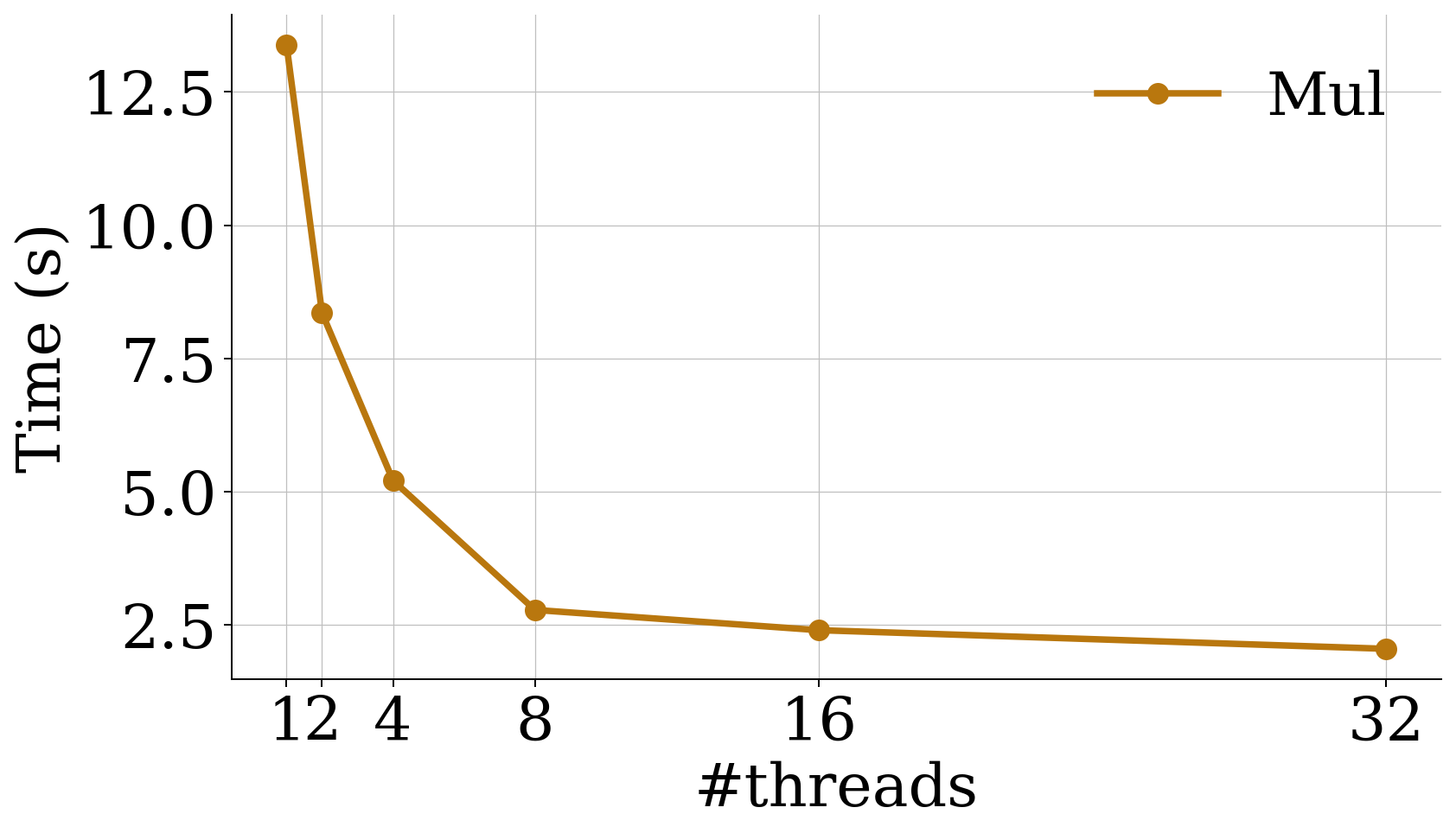}\vspace{-2mm}
		\subcaption{Multiplication}
		\label{fig:sorting-scalability-2pc}
	\end{minipage}
	\hfill
	\begin{minipage}[b]{0.24\textwidth}
		\centering
		\includegraphics[width=\textwidth]{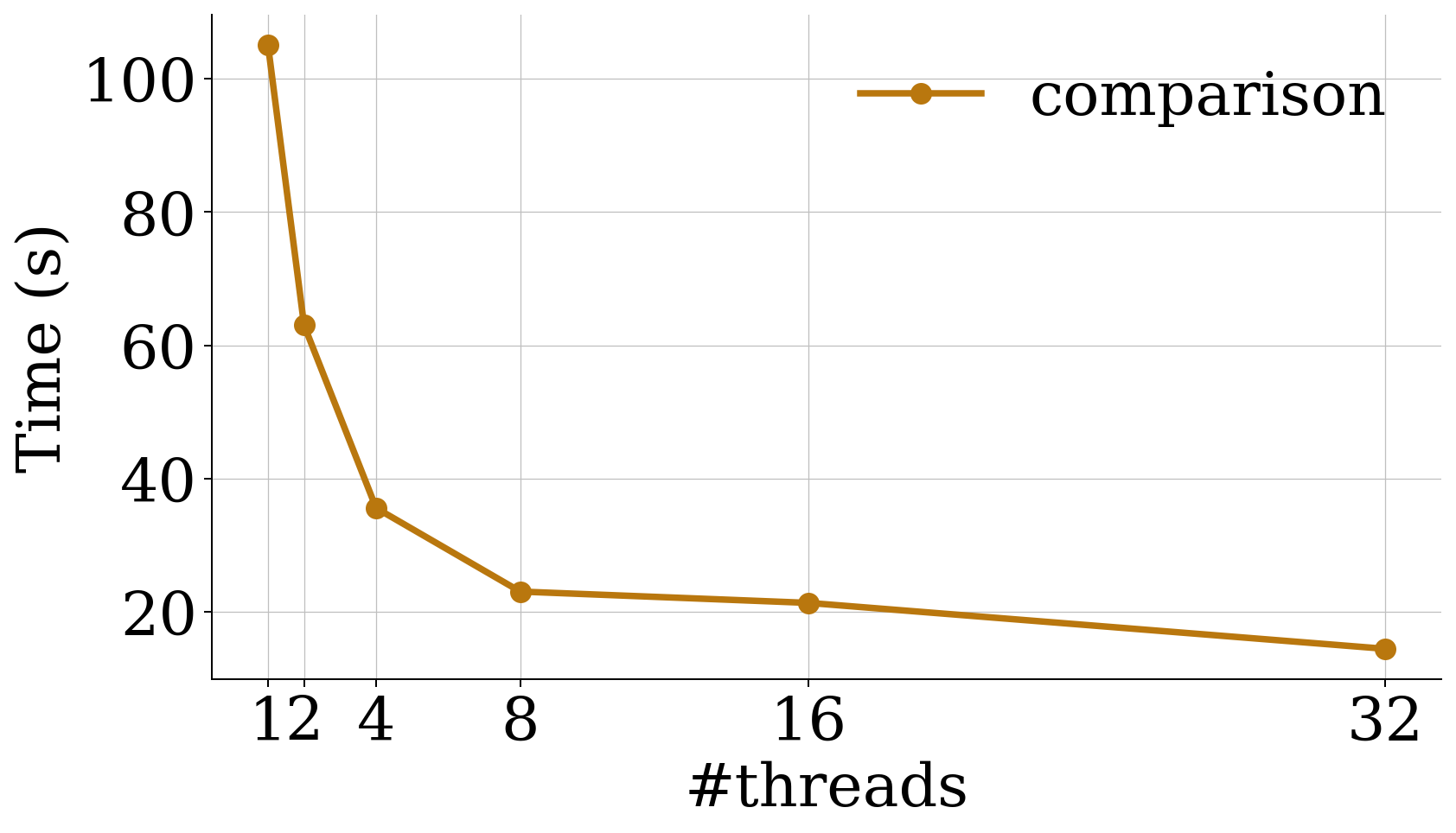}\vspace{-2mm}
		\subcaption{Comparison}
		\label{fig:gr-scalability-2pc}
	\end{minipage}
	\hfill
	\begin{minipage}[b]{0.24\textwidth}
		\centering
		\includegraphics[width=\textwidth]{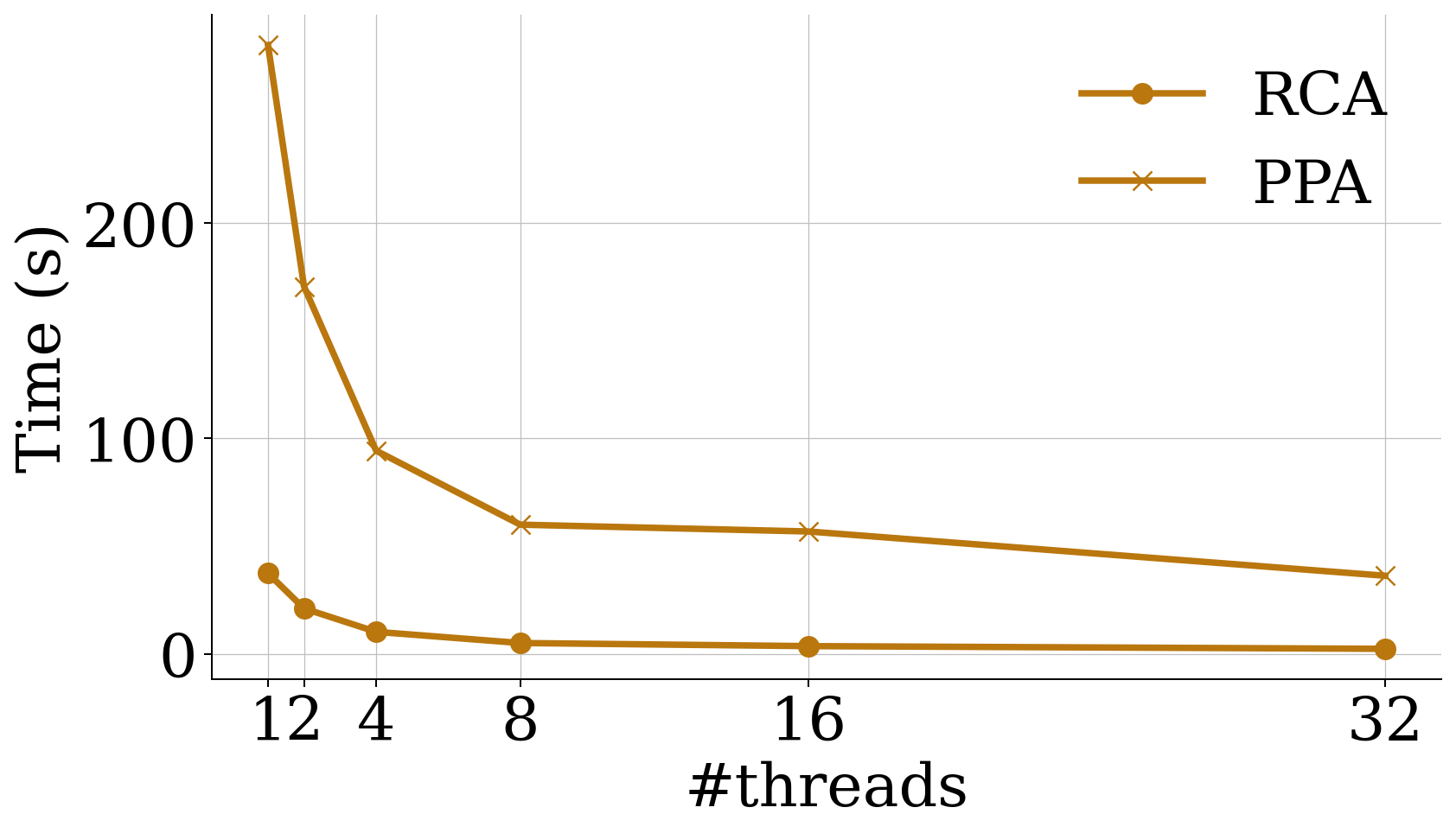}\vspace{-2mm}
		\subcaption{RCA/PPA}
		\label{fig:rca-ppa-scalability-2pc}
	\end{minipage}
	\hfill
	\begin{minipage}[b]{0.24\textwidth}
		\centering
		\includegraphics[width=\textwidth]{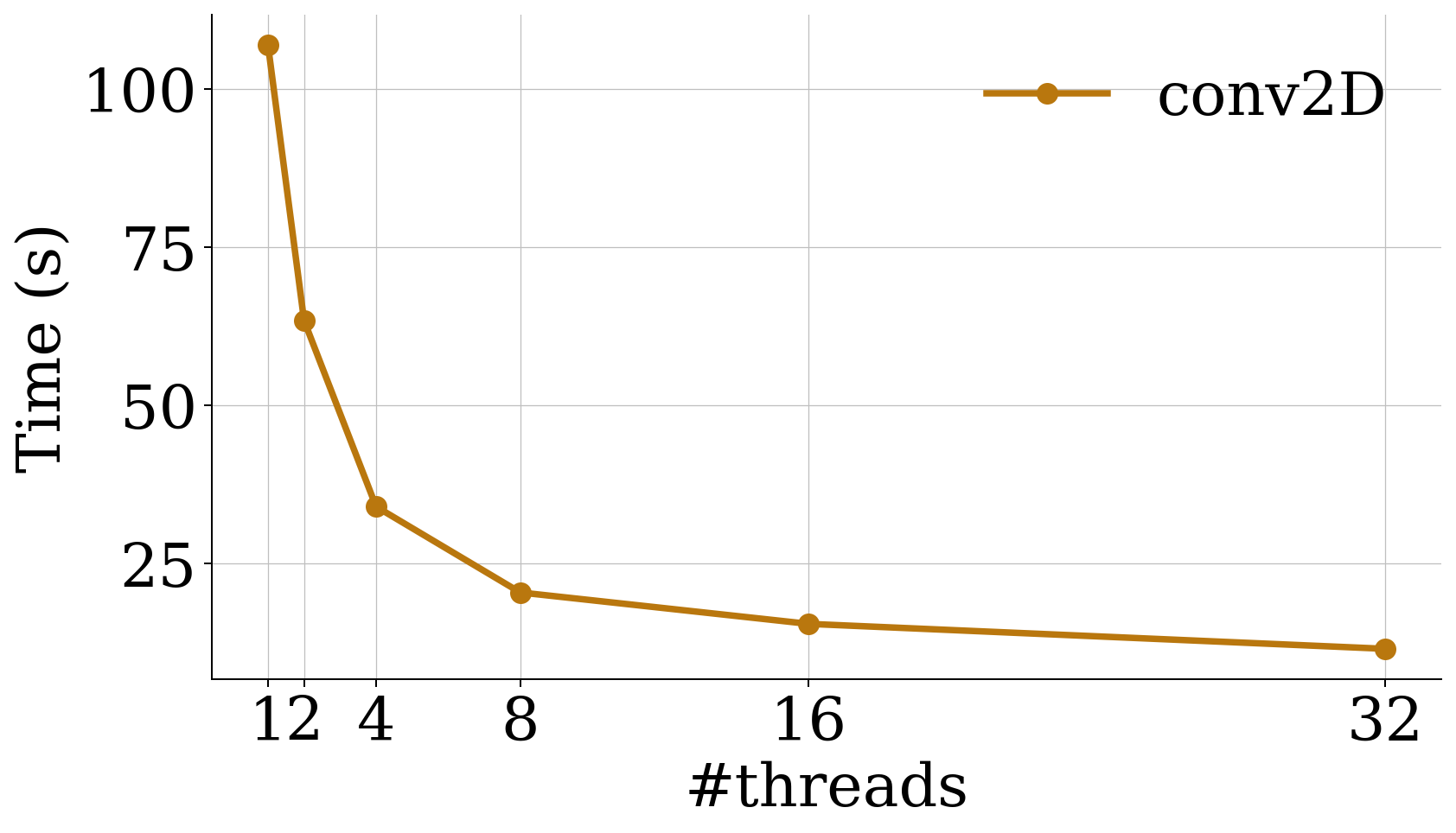}\vspace{-2mm}
		\subcaption{Conv2D}
		\label{fig:conv2d-scalability-2pc}
	\end{minipage}

	\vspace{-3mm}
	\caption{\ours's performance as we increase the number of worker threads per party~(ABY, \texttt{bm-LAN}).}\label{fig:scalability-2pc}
\end{figure*}

\begin{figure*}[t]
	\centering
	\begin{minipage}[b]{0.24\textwidth}
		\centering
		\includegraphics[width=\textwidth]{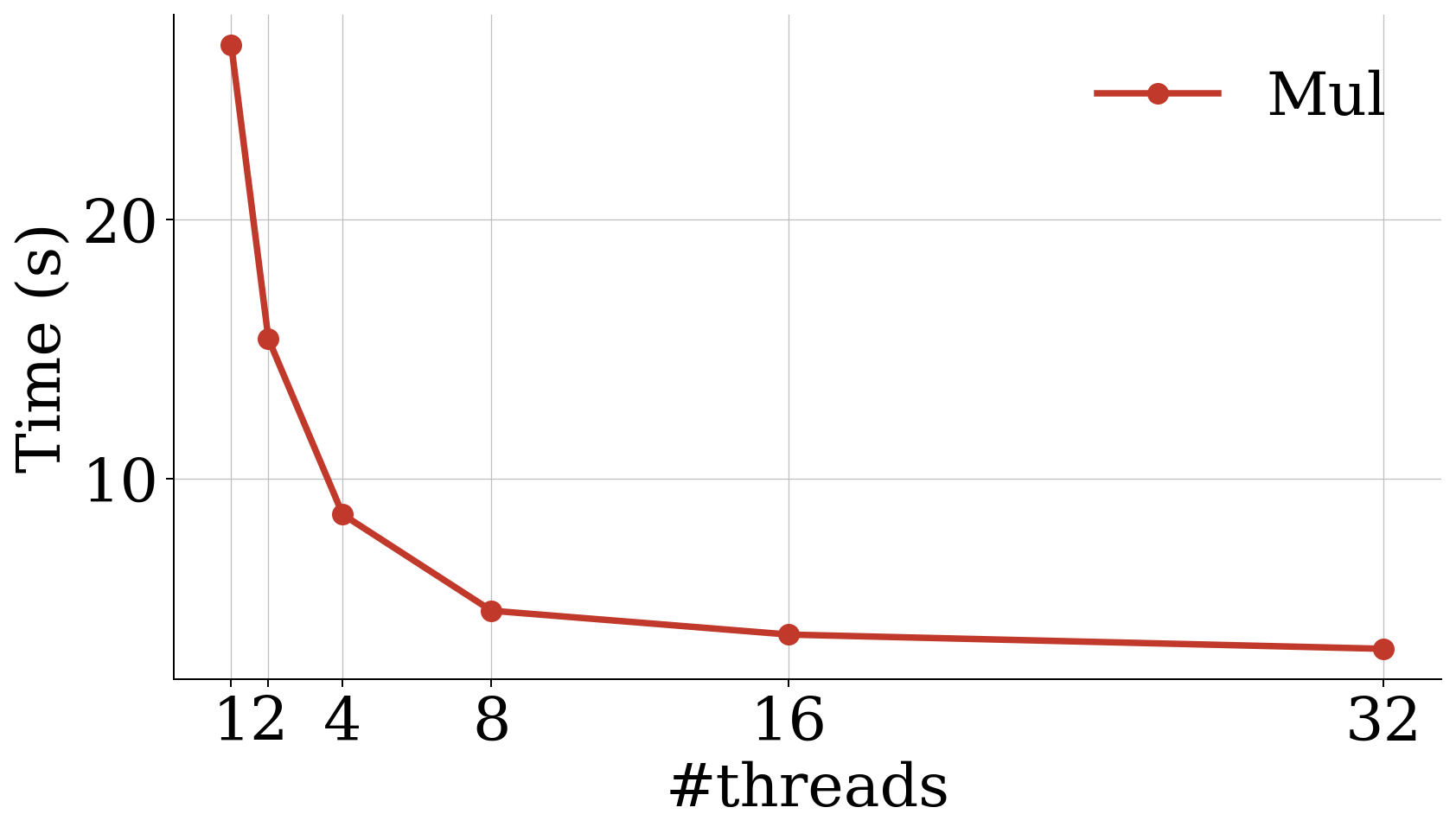}\vspace{-2mm}
		\subcaption{Multiplication}
		\label{fig:sorting-scalability-4pc}
	\end{minipage}
	\hfill
	\begin{minipage}[b]{0.24\textwidth}
		\centering
		\includegraphics[width=\textwidth]{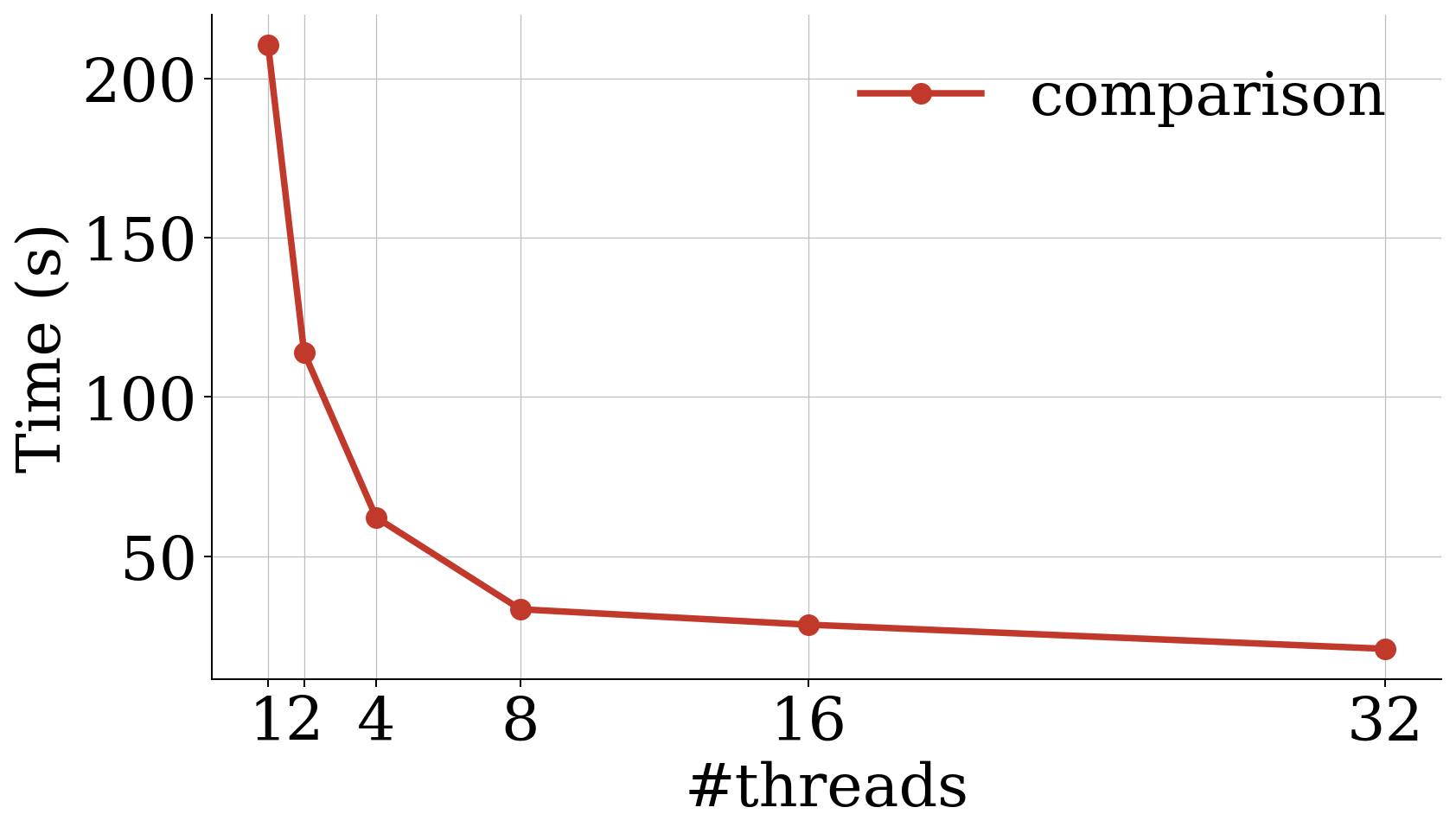}\vspace{-2mm}
		\subcaption{Comparison}
		\label{fig:gr-scalability-4pc}
	\end{minipage}
	\hfill
	\begin{minipage}[b]{0.24\textwidth}
		\centering
		\includegraphics[width=\textwidth]{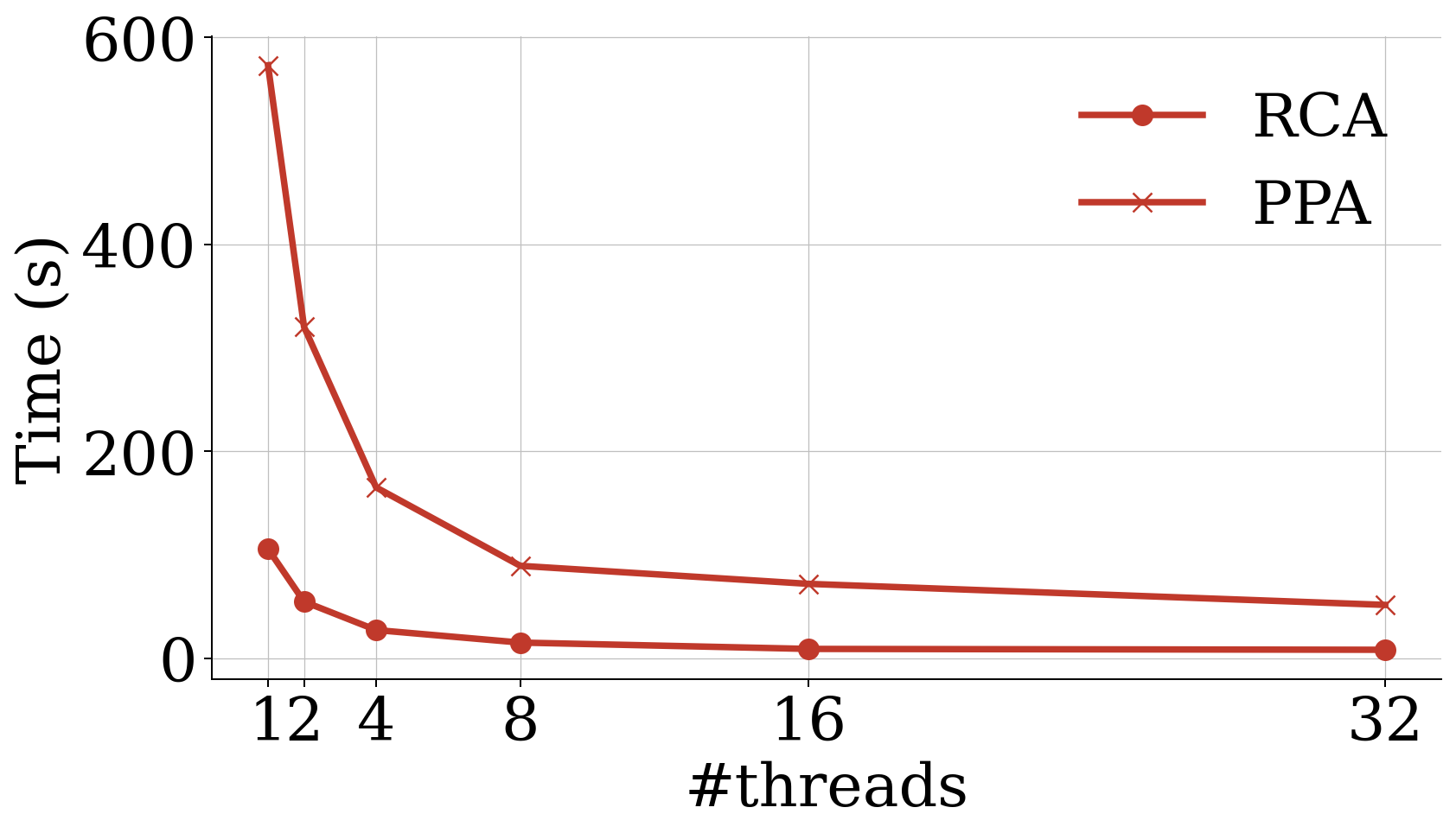}\vspace{-2mm}
		\subcaption{RCA/PPA}
		\label{fig:rca-ppa-scalability-4pc}
	\end{minipage}
	\hfill
	\begin{minipage}[b]{0.24\textwidth}
		\centering
		\includegraphics[width=\textwidth]{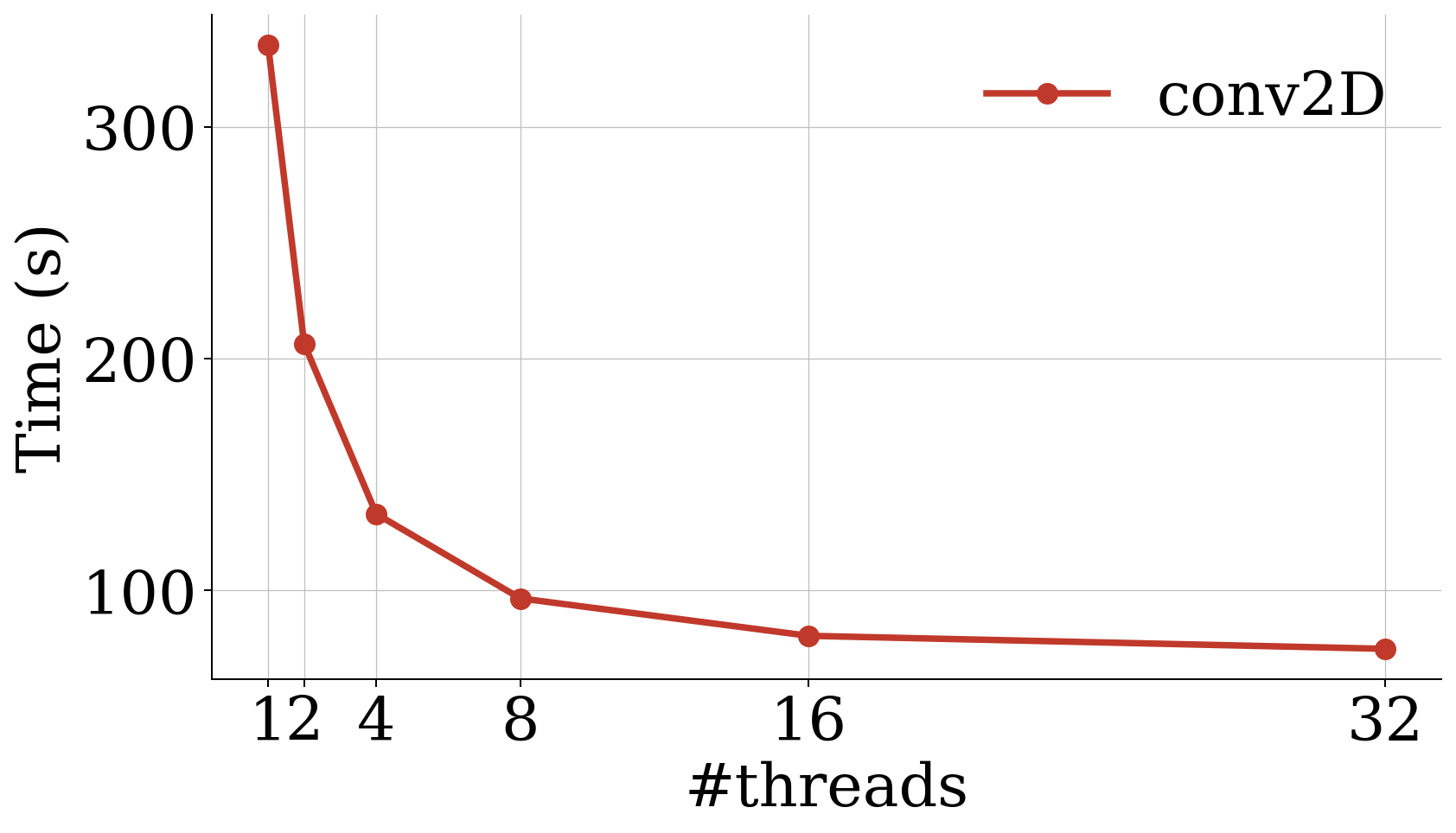}\vspace{-2mm}
		\subcaption{Conv2D}
		\label{fig:conv2d-scalability-4pc}
	\end{minipage}

	\vspace{-3mm}
	\caption{\ours's performance as we increase the number of worker threads per party~(Fantastic Four, \texttt{bm-LAN}).}\label{fig:scalability-4pc}
\end{figure*}

\begin{figure*}[t]
	\centering
	\begin{minipage}[b]{0.24\textwidth}
		\centering
		\includegraphics[width=\textwidth]{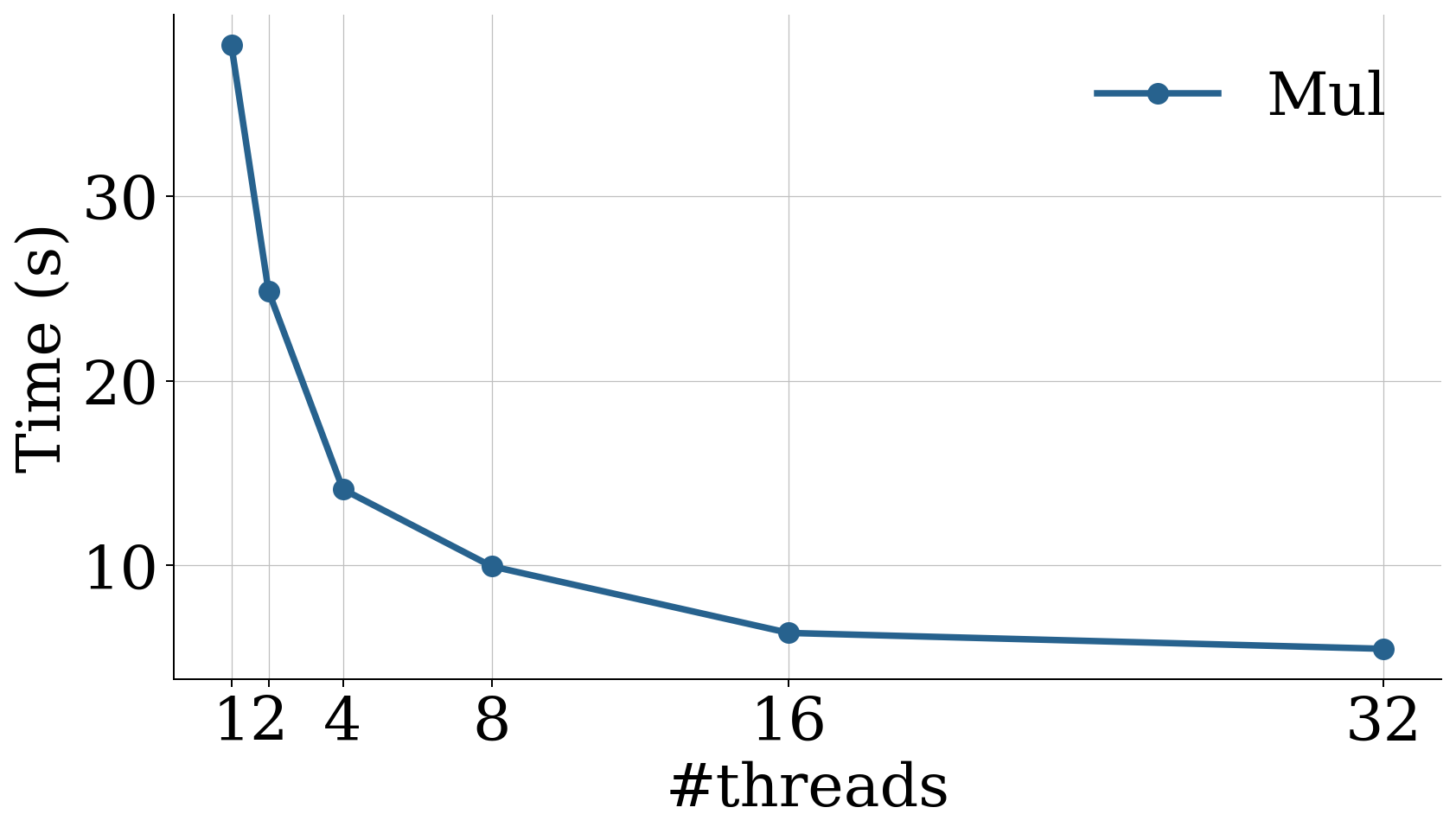}\vspace{-2mm}
		\subcaption{Multiplication}
		\label{fig:sorting-scalability-spdz}
	\end{minipage}
	\hfill
	\begin{minipage}[b]{0.24\textwidth}
		\centering
		\includegraphics[width=\textwidth]{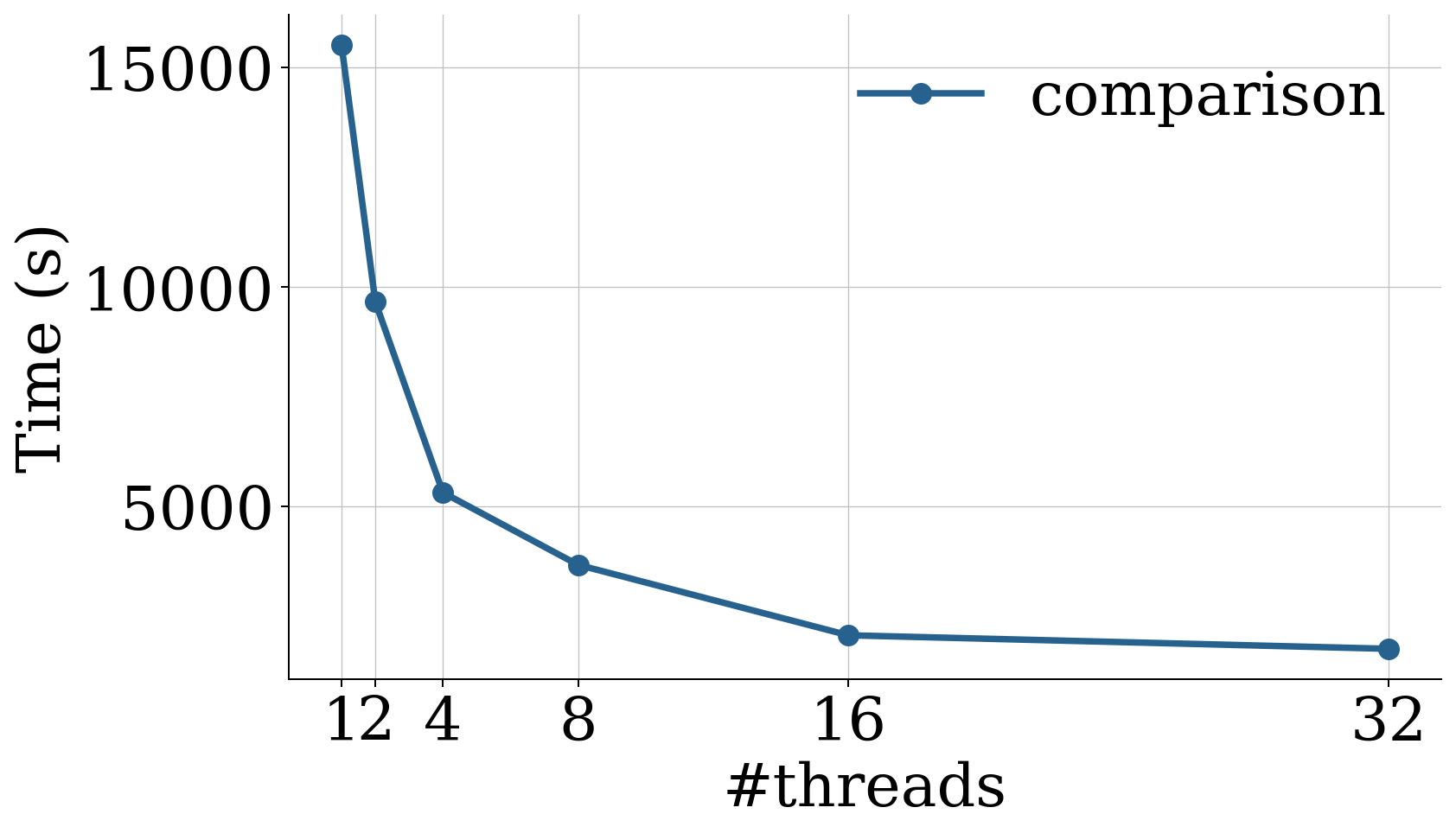}\vspace{-2mm}
		\subcaption{Comparison}
		\label{fig:gr-scalability-spdz}
	\end{minipage}
	\hfill
	\begin{minipage}[b]{0.24\textwidth}
		\centering
		\includegraphics[width=\textwidth]{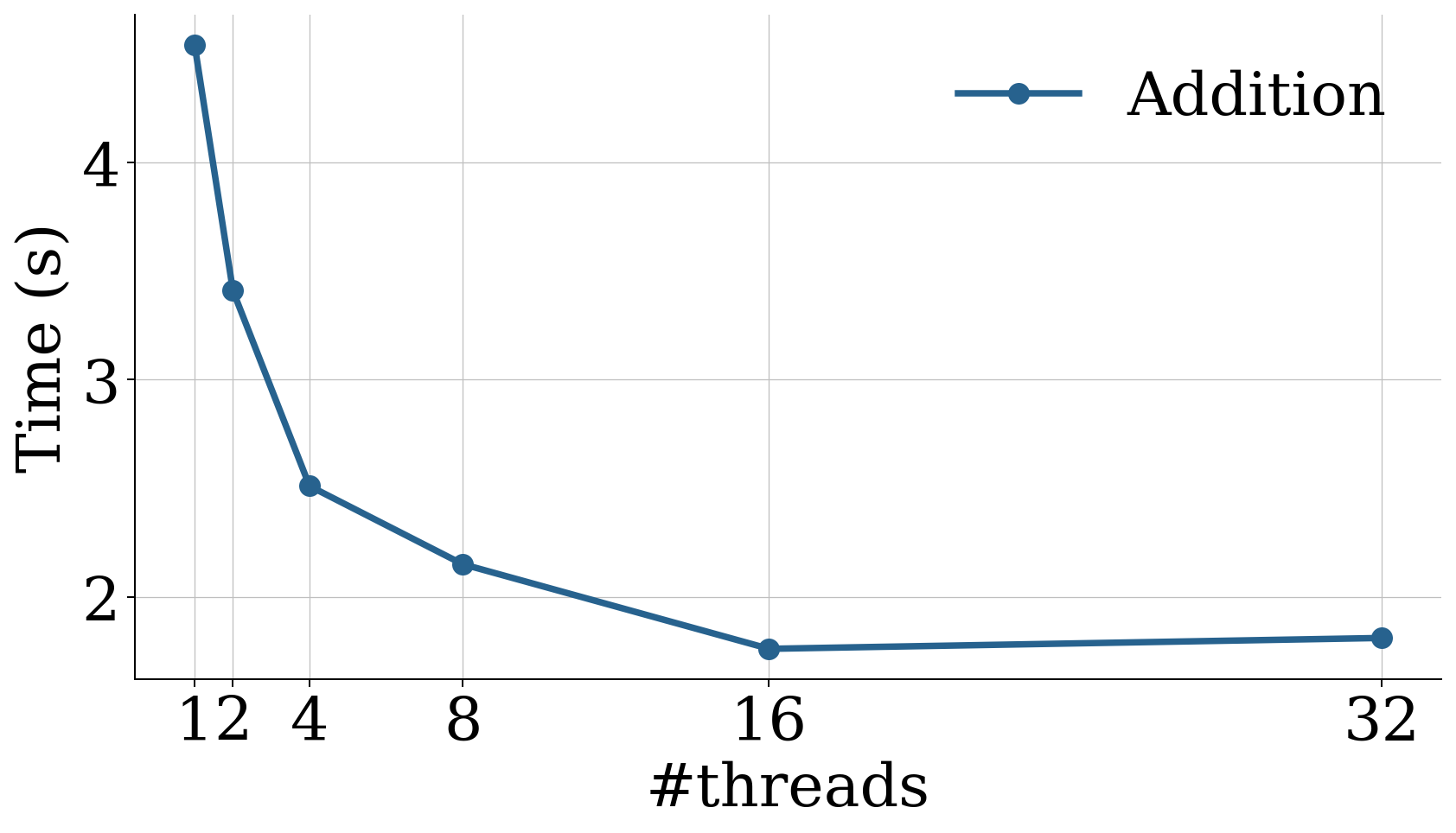}\vspace{-2mm}
		\subcaption{Addition}
		\label{fig:rca-ppa-scalability-spdz}
	\end{minipage}
	\hfill
	\begin{minipage}[b]{0.24\textwidth}
		\centering
		\includegraphics[width=\textwidth]{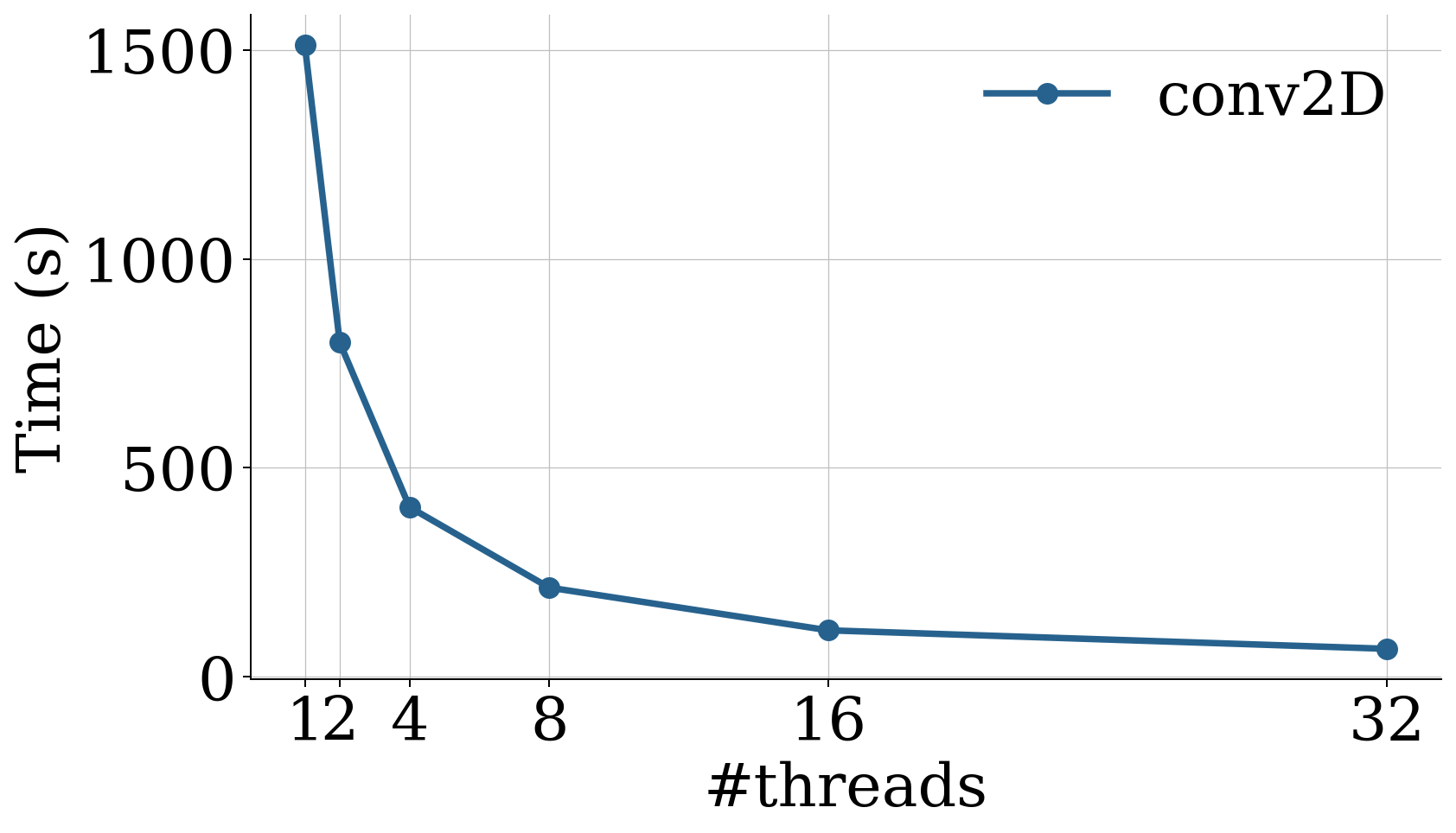}\vspace{-2mm}
		\subcaption{Conv2D}
		\label{fig:conv2d-scalability-spdz}
	\end{minipage}

	\vspace{-3mm}
	\caption{\ours's performance as we increase the number of worker threads per party~(SPDZ2k, \texttt{bm-LAN}).}\label{fig:scalability-spdz}
\end{figure*}

We evaluate the scalability of \ours's secure primitives for the ABY, Fantastic Four (F4), and SPDZ2k protocols. Results for ABY3 are given in the main paper (Fig.~\ref{fig:scalability}). We use vectors with $2^{28}$ 64-bit elements and run multiplication, comparison ($\geq$), boolean addition, and convolution in \texttt{bm-LAN}. Figures~\ref{fig:scalability-2pc}, \ref{fig:scalability-4pc}, and \ref{fig:scalability-spdz} show the execution times as we increase the number of threads per party from 1~to~32. For SPDZ2k, we never use RCA/PPA on boolean shares; therefore, we report performance for addition of arithmetic shares.

\subsection{Throughput of core primitives}
\camera{Table~\ref{tab:primitives-throughput} shows the throughput of \ours's core secure primitives: addition/XOR, multiplication/AND, and comparison ($\geq$), for all MPC protocols supported by the system. We measure throughput in \texttt{bm-LAN} and \texttt{bm-WAN} using 32 worker threads per party and vectors with $2^{28}$ 64-bit elements. Reported numbers are millions of operations per second.}

\begin{table}
\centering
\small
\setlength{\tabcolsep}{4pt}
\renewcommand{\arraystretch}{1.15}
\resizebox{\columnwidth}{!}{%
\begin{tabular}{l|cccc|cccc}
\hline
\multirow{2}{*}{\bf Primitive} & \multicolumn{4}{c|}{{\bf LAN}} & \multicolumn{4}{c}{{\bf WAN}} \\
\cline{2-9}
 & {\bf ABY} & {\bf ABY3} & {\bf F4} & {\bf SPDZ2k} & {\bf ABY} & {\bf ABY3} & {\bf F4} & {\bf SPDZ2k} \\
\hline
Addition / XOR       & 627   & 305.7 & 190.4 & 141.0 & 466.1 & 253.0 & 171.7 & 127.0 \\
Multiplication / AND & 111.8 & 167.6 & 67.4  & 42.27 & 14.9  & 19.19 & 12.8  & 9.9   \\
Comparison           & 12.6  & 29.8  & 9.42  & 0.13  & 4.0   & 7.0   & 3.1   & 0.072 \\
\hline
\end{tabular}%
}
\caption{\camera{Throughput of \ours's secure primitives in millions of operations per second for the ABY, ABY3, Fantastic Four (F4), and SPDZ2k protocols.}}
\label{tab:primitives-throughput}
\end{table}

\end{document}